\documentstyle[sprocl,latexsym,epsf]{article}

\input{psfig}

\def\be{\begin{equation}}
\def\ee{\end{equation}}
\def\bea{\begin{eqnarray}}
\def\eea{\end{eqnarray}}

\newcommand{\csw}{c_{\mbox{sw}}}

\begin{document}
{\hfill JLAB-THY-00-16}
\title{LATTICE GAUGE THEORY - QCD FROM QUARKS TO
HADRONS\footnote{Lectures given at the 14th.\ Annual Hampton
University Graduate Studies at CEBAF, 1st.\ to 18th.\ June, 1999}}
\author{D.\ G.\ RICHARDS}

\address{Jefferson Laboratory, MS 12H2, 12000 Jefferson
Avenue, Newport News,\\ VA 23606, USA\\[1.0ex]
Department of Physics, Old Dominion University, Norfolk,\\VA 23529,
USA\\E-mail: dgr@jlab.org}


\maketitle\abstracts{Lattice Gauge Theory enables an
\textit{ab initio} study of the low-energy properties of Quantum
Chromodynamics, the theory of the strong interaction.  I begin these
lectures by presenting the lattice formulation of QCD, and then outline
the benchmark calculation of lattice QCD, the light-hadron spectrum. I
then proceed to explore the predictive power of lattice QCD, in
particular as it pertains to hadronic physics.  I will discuss the
spectrum of glueballs, exotics and excited states, before investigating
the study of form factors and structure functions.  I will conclude by
showing how lattice QCD can be used to study multi-hadron systems, and
in particular provide insight into the nucleon-nucleon interaction.}

\section{Lattice QCD: the Basics}
\subsection{Introduction}\label{subsec:intro}
The fundamental forces of nature can by characterised by the strength
of the interaction: gravity, the weak interaction, responsible for
$\beta$-decay, the electromagnetic interaction, and finally the strong
nuclear force.  All but the weakest of these, gravity, are incorporated
in the Standard Model of particle interactions.

The Standard Model describes interactions through gauge
theories, characterised by a local symmetry, or gauge invariance.  The
simplest is the electromagnetic interaction, with the Abelian symmetry
of the gauge group U(1).  The model of Glashow, Weinberg and Salam
unified the electromagnetic and weak interactions through the a broken
symmetry group $\mbox{SU(2)}\otimes\mbox{U(1)}$.  The strong
interaction is associated with the unbroken non-Abelian symmetry group
SU(3), and is accorded the name Quantum Chromodynamics (QCD).

The strength of the electromagnetic interaction is characterised by the
dimensionless fine-structure constant $\alpha_e \simeq 1/137$.  A very
powerful calculational technique is to expand as a series in
$\alpha_e$ - perturbation theory.  QCD is characterised by a strong
coupling constant $\alpha_s \simeq {\cal O}(1)$.  QCD, however, is
asymptotically free, with an effective running coupling
$\alpha_s(Q^2)$ decreasing logarithmically with increasing $Q^2$.
Thus processes with an energy scale large compared with the natural
scale of the strong interaction, of the order of the proton mass, are often
amenable to the techniques of perturbation theory.

At energy scales of the order of the proton mass, perturbation theory
fails.  Yet a quantitative understanding of QCD is crucial both
for the study of the strong interaction, and for the study of the
other forces which are masked by the strong interaction.  In this
energy regime, we can either employ low-energy effective models of
QCD, or seek some way of performing a quantitative calculation
directly within QCD.  Lattice QCD is the only means we have of
performing such an \textit{ab initio} calculation.

Before proceeding to a description of lattice QCD, it is useful to
make a comparison between the properties of QCD and of QED:\\[1.0ex]
\begin{center}
\begin{tabular}{cccc}
& {\bfseries QED} & \textit{vs.} &{\bfseries
QCD}\\[1.0ex] 
\textit{Gauge particle} & Photon, ${\gamma}$ & &
Gluon, ${G}$ \\ 
\textit{Coupling to} & Electric charge, ${Q}$ & &
Colour charge \\ 
\textit{Charged particles} & $e, \mu, u, d, s\dots$ & & Quarks, $u,
d, s..., G$ \\
& \textit{Photon is neutral} & &
\textit{Gluon has colour charge} \\
\end{tabular}\\[1.0ex]
\end{center}
The gluon self-coupling reflects the non-Abelian, and highly
non-linear, nature of QCD.  Where are the quarks?  They are bound into
the colour-singlet hadrons we observe in nature.  Lattice gauge theory
provides the means to relate the quark degrees of freedom with the
observed hadronic degrees of freedom.

\subsection{Lattice Gauge Theory}\label{subsect:lgt}
Lattice gauge theory was proposed by Ken Wilson in
1974.\cite{wilson74}  Because of the gluon self-coupling, we have a
sensible pure-gauge theory of interacting gluons, even without quark, or
matter, fields.  We will consider this theory first.

We begin by formulating QCD in Euclidean space, which we accomplish by
a Wick rotation from Minkowski space,
\begin{equation}
t \rightarrow \tau \equiv it.
\end{equation}
The gauge fields are defined through
\begin{equation}
A_{\mu}(x) = A_{\mu}^a(x) T^a,
\end{equation}
where the $T^a, a = 1,\dots,8$ are the generators of SU(3), satisfying
\begin{eqnarray}
 [ T^a,  T^b ] & = & i f^{ab}_c\, 
 T^c \\
\mbox{Tr}\, T^a T^b & = &\frac{1}{2} \delta_{ab}.
\end{eqnarray}
We now introduce the field-strength tensor
\begin{equation}
F^{a}_{\mu\nu} \equiv \partial_{\mu} A^{a}_{\nu} - \partial_{\nu}
A^{a}_{\mu} + g f^{abc} A^{b}_{\mu} A^{c}_{\nu},
\end{equation}
in terms of which the Euclidean continuum action is
\begin{equation}
S = \frac{1}{4} \int d^4 x \, sF^a_{\mu\nu}
F^a_{\mu\nu}.\label{eq:action_continuum} 
\end{equation}
As we will see later, the crucial property of Euclidean space QCD for the
formulation of lattice gauge theories is that the action is real.
Gauge invariance is manifest through invariance under the
transformation
\begin{equation}
A_{\mu}(x) \to \Lambda(x) A_{\mu}(x) \Lambda^{-1}(x) - \frac{1}{ig}
(\partial_{\mu} \Lambda(x)) \Lambda^{-1}(x).\label{eq:gauge_trans}
\end{equation}

\begin{figure}
\begin{center}
\epsfxsize=250pt\epsfbox{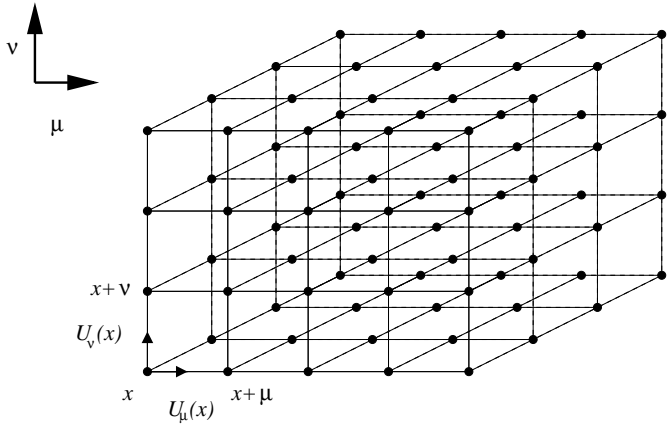}
\end{center}
\caption{A schematic of a lattice showing the association of the SU(3)
matrices $U_{\mu}(x)$ with the links of the lattice.\label{fig:grid}}
\end{figure}

We proceed to the lattice formulation of QCD by replacing a finite
region of continuum space-time by a discrete four-dimensional lattice,
or grid, of points.  The gluon degrees of freedom are represented by
SU(3) matrices $U_{\mu}(x)$ associated with the links connecting the
grid points, as shown in Figure~\ref{fig:grid}.  We work with the
elements of the group, rather than elements of the algebra, and the
SU(3) matrices $U_{\mu}(x)$ are related to the usual continuum gauge
fields through
\begin{equation}
U_{\mu}(x) = \exp i g \,  a \int_0^1 dt\,
A_{\mu}(x + t a \hat{\mu}),\label{eq:link_variable}
\end{equation}
where $g$ is the coupling constant, and $a$ the lattice spacing.
Under a gauge transformation $\Lambda(x)$, the link variables
transform as
\begin{equation}
U_{\mu}(x) \rightarrow \Lambda(x) U_{\mu}(x + \hat{\mu})
\Lambda^{-1}(x),
\end{equation}
in analogy with Eq.~\ref{eq:gauge_trans}.
Wilson's form of the lattice gauge action is constructed from the elementary
plaquettes~\cite{wilson74}
\begin{equation}
U_{\Box_{\mu\nu}}(x) = U_{\mu}(x) U_{\nu}(x
+ \hat{\mu}) U_{\mu}^{\dagger}(x + \hat{\nu}) U_{\nu}^{\dagger}(x).
\end{equation}
The plaquettes are clearly gauge invariant, and the action is then written
\begin{equation}
S_G  = \frac{2 N_c}{g^2} \sum_x \sum_{\mu > \nu} \left[ 1 -
\frac{1}{N_c} \Re \mbox{Tr}\, U_{\Box_{\mu\nu}}(x) \right]
 \equiv  - \frac{\beta}{N_c} \sum_x \sum_{\mu > \nu} \Re\,\mbox{Tr}
\, U_{\Box_{\mu\nu}},\label{eq:gauge_action}
\end{equation}
where we have ignored the constant term, and introduced
\[
\beta = \frac{2 N_c}{g^2}
\]
with, for QCD, $N_c = 3$.  It is straightforward to show that the
Wilson lattice gauge action is related to the continuum counterpart,
Eq.~(\ref{eq:action_continuum}), by
\begin{equation}
S_G = \frac{1}{4} \int d^4x \, F^a_{\mu\nu} F^a_{\mu\nu} + {\cal O}(a^2),
\end{equation}
so that the lattice gauge action has ${\cal O}(a^2)$ discretisation
errors.

\subsection{Observables and Lattice Gauge Simulations}
Within lattice gauge theory, the expectation value of an observable $O$
is given by the path integral
\begin{equation}
\langle O \rangle = \frac{1}{Z} \int {\cal D} U \, O(U) e^{-S_G(U)}
\end{equation}
where
\begin{equation}
{\cal D} U = \prod_{x,\mu} d U_{\mu}(x)
\end{equation}
and $Z$ is the generating functional
\begin{equation}
Z = \int {\cal D} U \, e^{-S_G(U)}; \, {\cal D} U = \prod_{x,\mu} d U_{\mu}(x).
\end{equation}
Before proceeding further, we need to define what we mean by the
integration over a group variable $d U$.  We do this through the Haar
measure, which for a compact group is the unique measure having the
following properties:
\begin{enumerate}
\item 
\[
\int_G dU \, f(U) = \int_G dU \, f(VU) = \int_G dU \, f(UV) \quad \forall\, V
\in G.
\]
\item 
\[
\int_G dU = 1.
\]
\end{enumerate}
This choice of measure respects gauge invariance.  Note that, because
we are employing the compact variables, $U_{\mu}(x)$, rather than the
elements of the algebra, we do not need to fix the gauge, and indeed
in most circumstances we do not do so.  However, there are cases
where working in a fixed gauge is useful, most notably in lattice
perturbation theory, where gauge fixing is essential, in the
definition of hadronic wave functions, where it is often useful to
work in Coulomb gauge, and most directly in the study of the
fundamental gluon and quark Green functions of the theory.

On a finite lattice, the calculation of observables is equivalent to
the evaluation of a very high, though finite, dimensional integral.
In principle, we could estimate this integral by evaluating the
integrand at uniformly distributed points.  This, however, would be
hopelessly inefficient; the exponential behaviour ensures that the
integral is dominated by regions where the action is small.  Instead
we use importance sampling, and generate gauge fields with a
probability distribution
\begin{equation}
e^{-S_G(U)}.
\end{equation}
The interpretation of this exponential in terms of a probability
distribution requires that the action be real, and hence the need to
work in Euclidean space.  The formulation follows that of many systems
in statistical physics.

\subsection{Statistical and Systematic Uncertainties}
\subsubsection{Statistical Uncertainties}
Observables in lattice QCD calculations arise from a Monte Carlo
procedure, and thus have statistical uncertainties.  Once we have
reached thermalisation, these uncertainties decrease as the square
root of the number of configurations, providing successive
configurations are sufficiently widely separated to be statistically
independent.  

\subsubsection{Systematic Uncertainties}
Of even greater delicacy than the statistical uncertainties are the
systematic uncertainties that enter our computations.  These arise from
a variety of sources, including:
\begin{itemize}
\item \textit{Finite Volume:} Our box must be sufficiently large that
finite volume effects are under control.  For light hadron
spectroscopy, box sizes of at least $2~\mbox{fm}$ are necessary to
ensure that the hadron is not ``squeezed'', but for excited states
even larger volumes may be required.  In addition, the requirement
that the spatial extent of the lattice be large compared with the
correlation length, set by the pseudoscalar mass, sets a still more
stringent constraint at the physical pion mass.
\item \textit{Discretisation Effects:} Increasing the inverse coupling
$\beta$ corresponds to progressing to weaker coupling, and hence
smaller lattice spacing $a$.  We must ensure that $\beta$ is
sufficiently large that the scale-breaking discretisation errors are
under control, and in practice we perform calculations at several
values of $a$ and extrapolate to the limit $a = 0$.
\end{itemize}
We will encounter several other potential sources of systematic errors
when we discuss the inclusion of the quarks.

\subsection{Including the Quarks}
The full generating functional for lattice QCD with a single flavour of
quark is
\begin{equation}
Z = \int {\cal D} U \, {\cal D}\psi \, {\cal D}
 \overline{\psi}
e^{-S_G(U) + \sum_{x,y} 
\overline{\psi} (x) M(x,y,U) \psi(y)},
\end{equation}
where $M(x,y,U)$ is the fermion
matrix which, in its ``na\"{i}ve'' form, is
\begin{equation}
M(x,y,U) = m\, \delta_{x,y} + \frac{1}{2} \sum_{\mu} \gamma_{\mu} \left(
U_{\mu}(x) \delta_{y, x+\hat{\mu}} - U_{\mu}^{\dagger}(x -
\hat{\mu}) \delta_{y, x - \hat{\mu}} \right)
\end{equation}
with $m$ the quark mass.  Because the fermion fields are represented
by Grassman variables, we can integrate out the fermion degrees of
freedom, to obtain
\begin{equation}
Z = \int {\cal D} U \, \det M(U) \, e^{-S_G(U)}.\label{eq:full_qcd}
\end{equation}
The determinant fluctuates rapidly between configurations.  Thus it is
not sufficient for a Monte-Carlo procedure to generate configurations
with probability $e^{-S_G(U)}$, and only include the determinant in
the calculation of observables; $\det M(U)$ must be included in the
measure.  Furthermore, whilst multiplication by the fermion matrix $M$
involves only nearest neighbour communication, the evaluation of $\det
M(U)$ is essentially a global operation.  Thus $\det M(U)$ must be
re-evaluated every time we update even a single value link variable
$U_{\mu}(x)$.  The most efficient algorithms for the simulation of QCD
with dynamical quarks, such as the Hybrid Monte Carlo
algorithm,\cite{hmc} involve a non-local updating procedure.
Nevertheless, calculations with dynamical quarks are at least 1000
times as expensive as pure gauge calculations.

The computational overhead of completely including the quark degrees
of freedom has encouraged many calculations to be performed in the
quenched approximation to QCD, in which we set
\[
\det M \equiv 1,
\]
in Eq.~(\ref{eq:full_qcd}).  This corresponds to suppressing the
contribution of closed quark loops in the path integral.  There are
two justifications for this seemly radical approximation.  Firstly,
the phenomenological observation that the neglect of the quark loops
corresponds to the neglect of OZI-suppressed processes.  Secondly, the
quenched approximation emerges in the large $N_c$ limit of QCD, where
$N_c$ is the number of colours.

\subsection{Fermion doubling and chiral symmetry}
Unfortunately, the lattice formulation of fermions presents a further
challenge.  To illustrate how this arises, let us consider the
momentum-space fermion propagator
\begin{equation}
M_{xy}^{-1} = \int_0^{\frac{\pi}{a}} \frac{d^4p}{(2\pi)^4} \frac{e^{i p
\cdot (x-y)}}{m + i \sum_{\mu} \gamma_{\mu} \sin a p_{\mu}}.
\end{equation}
In the massless limit, the propagator has a pole not only at $p_{\mu}
= 0$, but also at the edges of the Brillouin zone $p_{\mu} = \pi/a$.
Thus in four dimensions, we have a theory with $2^4$ non-interacting,
equal mass fermions.  This situation is a consequence of
the Dirac equation being first order.  Historically, there have been
two solutions to this problem.
\pagebreak
\subsubsection{Wilson Fermions}
Wilson proposed the addition of a second derivative term, or
momentum-dependent mass term, to the action:
\begin{eqnarray}
\lefteqn{S_F^W = \sum_x \left\{ (m + 4r) \bar{\psi}(x) \psi(x)
-\right. }\nonumber\\ &
&\left. \frac{1}{2}\sum_{\mu}\left[\bar{\psi}(x)(r-\gamma_{\mu})U_{\mu}(x)
\psi(x + \hat{\mu}) + \bar{\psi}(x + \hat{\mu})(r + \gamma_{\mu})
U^{\dagger}_{\mu}(x) \psi(x) \right]\right\}.\label{eq:wilson_quark}
\end{eqnarray}
In the continuum limit, we find
\begin{equation}
S_F^W = \int d^4 x \, \bar{\psi}(x) ( D + m - \frac{arD^2}{2})\psi(x)
+ {\cal O}(a^2).
\end{equation}
The addition of the second derivatives lifts the mass of the unwanted
doublers, but at a price.  We have added ${\cal O}(a)$ discretisation
errors to the fermion matrix, and furthermore the additional term
explicitly breaks chiral symmetry at any non-zero value of the lattice
spacing, though it is important to remember that chiral symmetry is
restored in the continuum limit.  The breaking of chiral symmetry has
the unfortunate consequence that the fermion masses are subject to an
additive mass renormalisation. In simulations, it is conventional to
reparameterise the fermion matrix as
\begin{equation}
M_{x,y} = (m + 4 r) \left\{ \delta_{x,y} 1 - \kappa \times
\mbox{``hopping term''} \right\}
\end{equation}
where the hopping parameter 
\begin{equation}
\kappa \equiv \frac{1}{2(4r + m)}
\end{equation}
is now a tunable parameter, reflecting the additive quark-mass renormalisation.
The critical value of the hopping parameter, $\kappa_{\rm crit}$, is
that value for which the pion mass vanishes.

\subsubsection{Kogut-Susskind Fermions}
The second approach, due to Kogut and Susskind, regards the complete
loss of chiral symmetry at finite $a$ as too great a sacrifice, and
therefore aims to preserve a remnant of chiral symmetry whilst
reducing the flavour-doubling problem.  This formulation leads to four
flavours of quark, but the different flavour and spin components are
assembled from fields at the sixteen corners of a $2^4$ hypercube.
While the Kogut-Susskind formulation has been extensively employed in
the calculation of quantities where chiral symmetry is crucial, the
problematic flavour identification means that I will concentrate on
the results using the Wilson formulation in these lectures.

\subsubsection{Chiral Fermions and the Ginsparg-Wilson
Relation}\label{sec:chiral} Is it possible to construct a formulation
that does indeed possess a symmetry analogous to chiral symmetry at a
finite lattice spacing, whilst admitting the correct spectrum of quark
states?  Let us list four
properties desired of the free fermion action, which we write in the
form
\[
S_F = a^4 \sum_{x} \bar{\psi}(x) D(x,y) \psi(y)
\]
\begin{enumerate}
\item $D(x,y)$ is local\label{enum:local}
\item Far below the cut-off, $D(p) \simeq i \gamma_{\mu} p_{\mu} +
{\cal O}(p^2)$.\label{enum:cont}
\item $D(p)$ is invertible at all non-zero momenta.\label{enum:double}
\item $\gamma_5 D + D \gamma_5 = 0$.\label{enum:chiral}
\end{enumerate}
The last two requirements require explanation; \ref{enum:double}
demands that the only poles occur at zero momentum, and hence that
there be no doublers, whilst \ref{enum:chiral} is just a statement
of chiral symmetry.  The famous Nielsen-Ninomiya ``no-go''
theorem~\cite{nielsen81} states that it is not possible to find a Dirac
operator that allows all four requirements to be satisfied
simultaneously, and since the first two requirements were deemed
sacrosanct either chiral symmetry had to be broken, or
flavour-doubling had to be accepted.

The recent revolution in our understanding of chiral fermions are related
to the rediscovery of the Ginsparg-Wilson relation~\cite{ginsparg82}
\begin{equation}
\gamma_5 D + D \gamma_5 = a D \gamma_5 D,\label{eq:ginsparg}
\end{equation}
as a replacement for requirement~\ref{enum:chiral}.  At non-zero
distances, the Dirac operator does indeed commute with $\gamma_5$,
and a symmetry, reducing to chiral symmetry in the continuum limit, is
preserved even at non-zero values of the lattice spacing.  The problem
is finding a formulation of the Dirac operator that does indeed
satisfy the relation of Eq.~\ref{eq:ginsparg}.

Recently, two formulations satisfying this relation have been
discovered.  In the case of Domain Wall
fermions~\cite{kaplan92,shamir93} (DWF), an auxiliary fifth dimension,
with coordinate $s$ and extent $N_s$, is introduced.  The action is
essentially a five-dimensional Wilson fermion action
\begin{equation}
S_F^{\rm DWF} = \sum_{x,y,s,s'} \bar{\psi}(x) \left( D(x,y)
\delta_{s,s'} + D^5(s,s') \delta_{x,y} \right) \psi(y),\label{eq:dwf}
\end{equation}
where $D(x,y)$ is the usual Wilson-Dirac operator introduced in
Eq.~\ref{eq:wilson_quark}, but with a \textit{negative} mass term $M$.
The operator in the fifth dimension, $D^5(s,s')$, couples the boundaries
through a parameter $-m$, which is proportional to the usual
four-dimensional quark mass; note that no gauge links are introduced
in the fifth dimension.  The chiral limit corresponds to $L_s
\rightarrow \infty$, and then $m \rightarrow 0$; following the former
of these limits, the quark mass is only multiplicatively
renormalised.  

\begin{figure}
\begin{center}
\epsfxsize=250pt\epsfbox{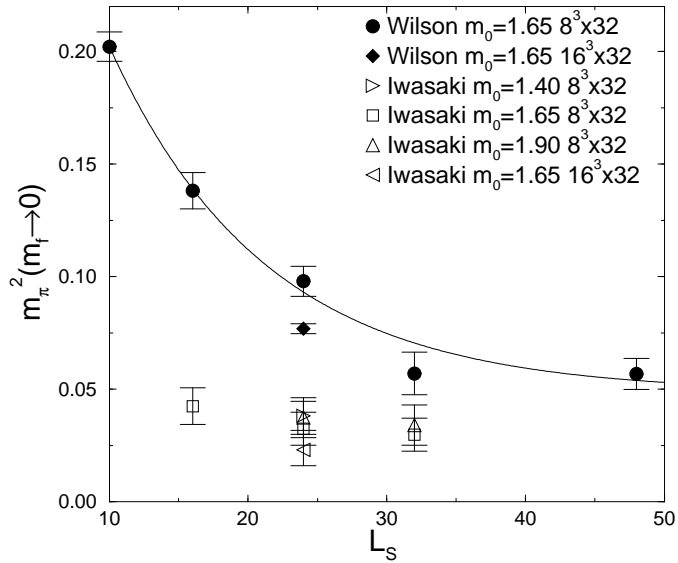}
\end{center}
\caption{The residual pion mass $m_{\pi}^2$ as a function of the
extent $L_s$ of the fifth dimension in the DWF formulation.\protect\cite{wu99}
The calculation is performed in the quenched approximation to QCD, and
\textit{Wilson} and \textit{Iwasaki} refer to the standard Wilson
gauge action, and an improved gauge action
respectively.}\label{fig:dwf_pion}
\end{figure}

A crucial issue is how small a value of $L_s$ is sufficient to
maintain good chiral properties whilst minimising the computational
cost.  For the case of hadronic physics, perhaps more important than
having good chiral properties is being able to perform simulations at
sufficiently small quark mass for the pion cloud to emerge.
Fig.~\ref{fig:dwf_pion} shows the residual pion mass in the chiral
limit as a function of function of $L_s$.\cite{wu99}

The second approach is the Overlap formalism, introduced by Narayanan
and Neuberger.\cite{narayanan93,narayanan95}  Here the overlap-Dirac
operator satisfying the property \ref{enum:chiral} is
\begin{equation}
D^N = \frac{1}{2} ( 1 + \mu + (1-\mu) \gamma_5
\frac{H(m)}{\sqrt{H(m) H(m)}}\label{eq:overlap}
\end{equation}
where $H(m)$ is the Hermitian Wilson-Dirac fermion operator, with
negative mass $m$, defined by $H = \gamma_5 D$ where $D$ is the usual
Wilson-Dirac operator.  The parameter $\mu$ is related to the
physical quark mass.  In this case, the extra computational cost comes
not from computing in five dimensions, but rather from evaluating the
step function
\begin{equation}
\epsilon(H) = \frac{H}{\sqrt{H H}}.
\end{equation}
The relative computational overheads of the two implementations is the
subject of intense investigation,\cite{edwards00} but in any case the
overhead compared to the standard Wilson fermion action is
considerable.  Whilst this overhead is justified for chiral gauge
theories, the situation in the case of hadronic physics is less clear;
perhaps there are more efficient ways of approaching physical values
of the light-quark masses.

\subsection{Improvement}
\begin{figure}
\begin{center}
\epsfxsize=250pt\epsfbox{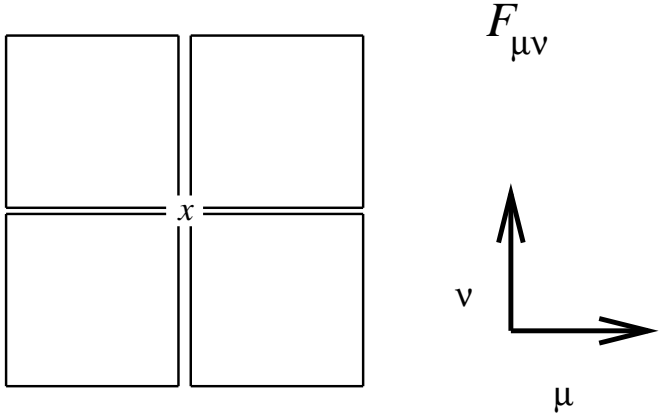}
\end{center}
\caption{The lattice discretisation of the field-strength $F_{\mu\nu}(x)$
in terms of the plaquettes with corners at $x$}\label{fig:clover}
\end{figure}

The addition of the Wilson term to the fermion action has introduced
${\cal O}(a)$ discretisation errors; in contrast the gauge action has
only ${\cal O}(a^2)$ discretisation errors.  Thus there has been an
emphasis on reducing the errors in the fermion sector through the
addition of higher dimensional terms to the action, the improvement
programme of Symanzik.\cite{symanzik83} In the case of the Wilson
fermion action, the leading ${\cal O}(a)$ errors can be removed
through the addition of a single dimension-five operator, the magnetic
moment, or clover term, proposed by Sheikholeslami and
Wohlert~\cite{SWaction}
\begin{equation}
S_F = S_F^W - i \frac{\csw \kappa}{2} \sum_{x,\mu,\nu}
\bar{\psi}(x) F_{\mu\nu}(x) \sigma_{\mu\nu} \psi(x).
\end{equation}
The name ``clover'' is clear from the natural lattice discretisation of
$F_{\mu\nu}$ illustrated in Figure~\ref{fig:clover}.

Using tree-level perturbation theory, the clover coefficient $\csw$ is
unity, and the discretisation errors on hadron masses and, with an
appropriate discretisation of operators, on-shell matrix elements are
formally ${\cal O}(a g^2)$.~\cite{heatlie91} UKQCD performed an
extensive investigation of the hadron spectrum and hadronic matrix
elements using this value of $\csw$, and the discretisation errors,
particularly for systems containing heavy quarks, can be
substantial.~\cite{hyperfine}

More recently, two prescriptions for determining $\csw$ have been
proposed with the aim of reducing discretisation errors still
further. In the first, the clover coefficient is constrained to its
mean-field-improved, or tadpole, value~\cite{lm93}
\begin{equation}
\csw = \mbox{TAD} = \frac{1}{u_0^3}\label{eq:csw_tad}
\end{equation}
where
\begin{equation}
u_0 = \langle \frac{1}{3} \mbox{Tr} U_{\Box} \rangle
\end{equation}
is an estimate of the mean-value of the link variable $U_{\mu}$.
Though formally the discretisation errors remain $O(a g^2)$, this
prescription recognises the poor behaviour of naive lattice
perturbation theory arising from the ``tadpole'' contributions, and
attempts to resum the dominant higher-order contributions through the
use of a more physical expansion parameter.

The second prescription~\cite{alpha96,luscher96} determines $\csw$
non-perturbatively in such a way as to remove \emph{all} ${ cal O}(a)$
discretisation errors from hadron masses, and, with an appropriate
choice of operators, from all on-shell matrix elements~\cite{alpha97}
\begin{equation}
\csw = \mbox{NP} = \frac{1 - 0.656 g_0^2 - 0.152 g_0^4 - 0.054
g_0^6}{1 - 0.922 g_0^2},\quad g_0^2 \le 1,\label{eq:csw_np}
\end{equation}
where $g_0^2 = 6/\beta$.  

We end this section by remarking that the chiral-fermion formulations
are already ${\cal O}(a)$-improved; the ${\cal O}(a)$ discretisation
are introduced through the chiral-symmetry-breaking Wilson term.

\section{The light-hadron spectrum}
The calculation of the spectrum of hadrons containing the light quarks
(u,d,s) is the benchmark calculation of lattice QCD; we know many of
the results!  It also provides a useful theatre for discussion of
some of the issues I raised in the introduction.  However, let us
begin with one of the simplest observations we can make in
lattice QCD, that of the linear confining potential.

\subsection{The Static Quark Potential and Quark Confinement}
\begin{figure}
\begin{center}
\epsfxsize=300pt\epsfbox{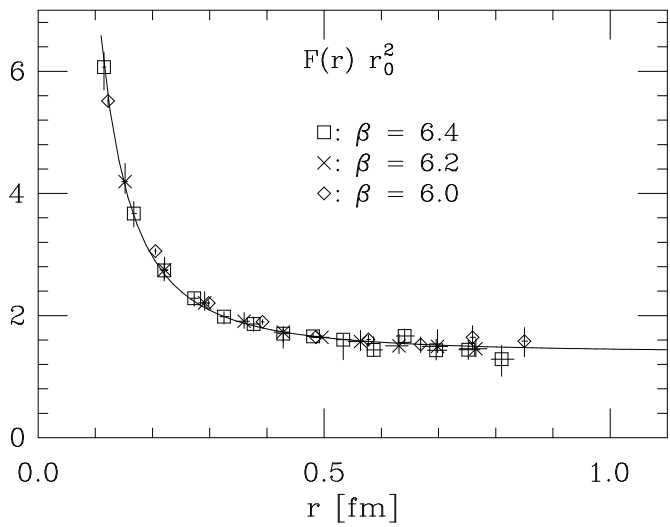}
\end{center}
\caption{The force between two static quarks, in units of Sommer's
$r_0$ parameter,\protect\cite{sommer94} is shown at three values
of the lattice spacing, corresponding to $\beta = 6.0, 6.2\mbox{~and~}
6.4$.\protect\cite{ukqcd94}}
\label{fig:ukqcd_potential}
\end{figure}

The simplest observable we can obtain from a lattice simulation is the
potential between two (infinitely heavy) static quarks.  We construct
the Wilson loops
\begin{equation}
W(R,T) = \langle \mbox{Tr}\, U(C(R,T))\rangle
\end{equation}
where $U(C(R,T))$ is the product of gauge links around a $R \times T$
space-time loop.
\begin{figure}
\begin{center}
\epsfxsize=200pt\epsfbox{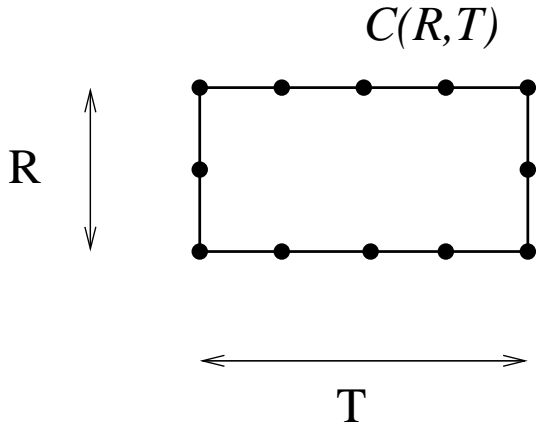}
\end{center}
\caption{The construction of the $R\times T$ wilson loop in a
space-time plane.}\label{fig:wilson_loop}
\end{figure}
At large times, we can extract the potential $V(R)$ between two static
quarks $Q$ at separation $R$, using
\begin{equation}
W(R,T) \stackrel{T \rightarrow \infty}{\sim} e^{ -T\, V(R)}.
\end{equation}
An area-law decay of the large Wilson loops is characteristic of a
linear, confining potential in QCD, giving rise to a constant force
with increasing separations $r$.  We parameterise this force $F(r)$ by
\begin{equation}
F(r) r_0^2 = 1.65 + \frac{\pi}{12}\left(\frac{r_0^2}{r^2} - 1 \right)
\end{equation}
where $r_0 \simeq 0.5~\mbox{fm}$ is a phenomenological
parameter~\cite{sommer94} which
we use to determine the scale.  We show this force for the
pure-gauge theory, corresponding to the quenched approximation, in
Figure~\ref{fig:ukqcd_potential}.  At large separations we see the
constant force indicative of confinement.  But there is a further
important observation. The results from all three calculations lie on
a single curve, even though the calculations span a factor of two in
lattice spacing $a$, from about $0.05~\mbox{fm}$ to $0.1~\mbox{fm}$.

\begin{figure}
\begin{center}
\epsfxsize=350pt\epsfbox{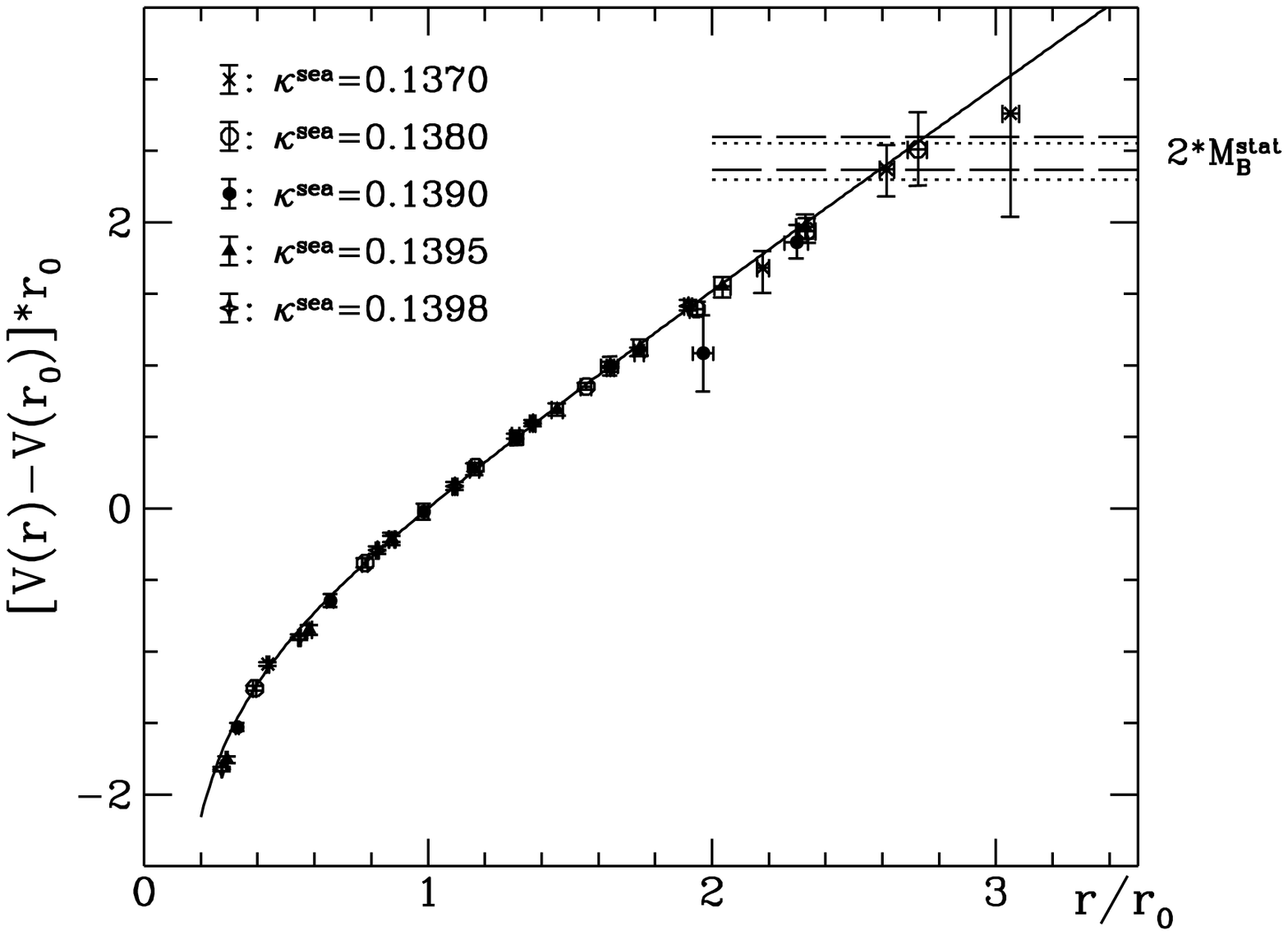}
\end{center}
\caption{The scaled and normalised potential as a function of $r/r_0$,
where $r_0$ is the Sommer scale discussed earlier, as obtained on a
$12^3\times 24$ lattice, using ${\cal O}(a)$-improved dynamical
fermions.~\protect\cite{ukqcd98}  The expected region of string
breaking is shown by the horizontal lines.}\label{fig:ukqcd_string}
\end{figure}

Do we expect the same picture of a rising, linear static-quark
potential in full QCD, with dynamical quarks?  As the two heavy quarks
are separated, the energy stored in the string increases.  Eventually,
the string can ``break'' to form a quark-antiquark pair, a process
that does not occur in the pure-gauge theory.  This should lead to a
flattening of the potential at some distance corresponding to an
energy in the flux tube of $2 M_B$, twice the binding energy of
static-heavy-light meson.  In practice, there has been no clear
observation of such a feature at zero temperature from the simple
Wilson loop operator; the behaviour of the potential in full QCD from
a calculation using an ${\cal O}(a)$-improved fermion
action~\cite{ukqcd98} is shown in Figure~\ref{fig:ukqcd_string}.

\begin{figure}
\begin{center}
\epsfxsize=250pt\epsfbox{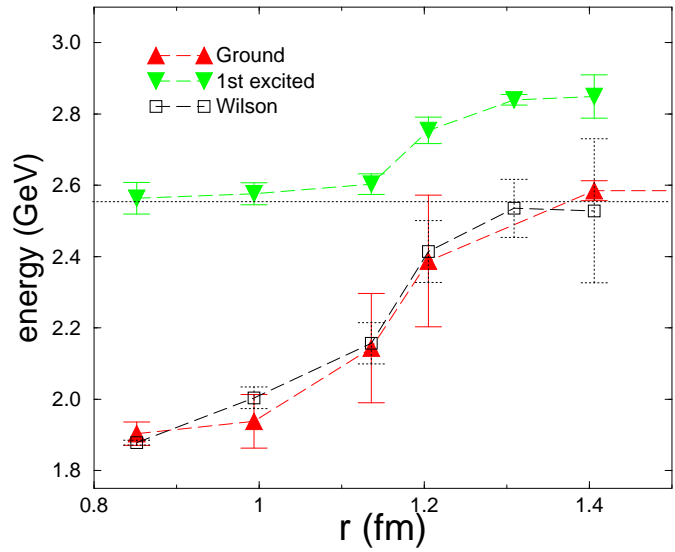}
\end{center}
\caption{The ground and excited state energies obtained from a
variational calculation including the $Q\bar{Q}$ and
$Q\bar{q}\bar{Q}q$ operators.\protect\cite{pennanen00} The horizontal
line is $2 M_B$, the static binding energy shown also
in Figure~\protect\ref{fig:ukqcd_string}}
\label{fig:sbgev}
\end{figure}

As the string breaks, the $Q\bar{Q}$ system crosses to a system of two
heavy-light $\bar{Q}q$ mesons.  The lack of observation of the string
breaking from the Wilson loop is ascribed to the poor overlap of the
Wilson loop operator with two such heavy-light mesons.  The mixing
between these two states has been successfully investigated in simpler
systems~\cite{philipsen98,knechtli98,stephenson99}, and recently a
study has been made within QCD.\cite{pennanen00} Both the ground state
and first excited state energies are extracted by a variational
calculation using the Wilson loop and $Q\bar{q}\bar{Q}q$ operators, as
shown in Figure~\ref{fig:sbgev}.  We will study a related problem when
we discuss the nucleon-nucleon interaction in
Section~\ref{sec:multihadron}.

\subsection{Spectrum recipe}\label{sec:spectrum_recipe}
The elemental building blocks of a
spectrum calculation are the quark propagators
\begin{equation}
G^{ij}_{\alpha\beta}(x,y) = \langle 0 | \psi^i_{\alpha}(x)
\bar{\psi}^j_{\beta} (y) | 0 \rangle.\label{eq:quark_prop}
\end{equation}
The quark propagator to every point $x$ on the lattice from a
\textit{fixed}
source point $y$, or linear combination of source points, is obtained
by inverting the fermion matrix for a fixed source vector.  There are
a variety of linear-solver methods used to accomplish this.

In principle, the recipe for determining the mass of a ground-state
hadron $P$ is straightforward:
\begin{enumerate}
\item Choose an interpolating operator ${\cal O}$ that has a good
overlap with $P$
\[
\langle 0\mid {\cal O} \mid P\rangle \ne 0,
\]
and ideally a small overlap with other states having the same quantum
numbers.
\item Form the time-sliced correlation function
\[
C(t) = \sum_{\vec{x}} \langle {\cal O}(\vec{x}, t) {\cal O}^{\dagger}
(\vec{0}, 0).
\]
This is expressed using a Wick expansion in terms of the quark propagators
of Eq.~\ref{eq:quark_prop}
\item  Insert a complete set of states between ${\cal O}$ and ${\cal
O}^{\dagger}$.  The time-sliced
sum puts the intermediate states at rest, and we find
\begin{eqnarray*}
C(t) & = & \sum_{\vec{x}} \sum_P \int \frac{d^3k}{(2 \pi)^3 2 E(\vec{k})}
\langle 0 |
{\cal O}(\vec{x}, t) | P(\vec{k}) \rangle \langle P(\vec{k}) | {\cal
O}^{\dagger} (\vec{0},0) | 0 \rangle\\
& = & \sum_P \frac{\mid \langle 0 \mid {\cal O} \mid P
\rangle \mid^2}{2 \, M_P} e^{i M_P t}.
\end{eqnarray*}
\item Continue to Euclidean space $t \rightarrow i t$, and we find
\begin{equation}
C(t) = \sum_P \frac{\mid \langle 0 \mid {\cal O} \mid P
\rangle \mid^2}{2 \, M_P} e^{- M_P t}.\label{eq:spectral_sum}
\end{equation}
\end{enumerate}
At large times, the lightest state dominates the spectral sum in
Eq.~(\ref{eq:spectral_sum}), and we can extract the ground state mass.

However there are many considerations that complicate this picture.
Firstly, the temporal separation between the hadrons must be
sufficiently large that the ground state can be identified in
Eq.~(\ref{eq:spectral_sum}); we aim to accomplish this by choosing an
operator having a large overlap with the ground state relative to the
excited states, and by fitting to several interpolating operators.
Secondly, the correlation lengths in our calculation must be small
compared with the size of the box in which we are working.  This
correlation length is simply the inverse of the pion mass, and we
require $m_{\pi} L \simeq 5$, where $L$ is the spatial extent of the
lattice.  Most, though not all, simulations to date have restricted
consideration to quark masses in the region of the strange-quark mass.

The benchmark calculation of the quenched light-hadron spectrum using
the unimproved fermion action has been performed by the CP-PACS
collaboration.\cite{cppacs2000} They perform their calculation on a
variety of lattice sizes, to control finite size effects, and at a
variety of lattice spacings, to enable an extrapolation to the
continuum limit.

Recently, there has been a similar calculation using
the non-perturbatively improved clover fermion action by the UKQCD
Collaboration.\cite{ukqcd2000}  They also perform an extrapolation to
the continuum limit, but in their case the discretisation
uncertainties are ${\cal O}(a)$ rather than ${\cal O}(a^2)$, and I
will describe some of the details of this calculation.

UKQCD generated $16^3\times32$ lattices at $\beta = 5.7$,
$16^3\times48$ lattices at $\beta = 6.0$ and $24^3\times48$ lattices
at $\beta = 6.2$, corresponding to approximately a factor two span of
lattice spacings but at roughly the same spatial volumes.  In
addition, a calculation was performed on a larger $32^3\times64$
lattice at $\beta = 6.0$ to enable a study of finite-volume effects.
Quark propagators were computed with the clover coefficient having
both its tadpole-improved value $\csw = \mbox{TAD}$ ($\beta = 5.7,
6.0,6.2$), and its non-perturbatively determined value $\csw=
\mbox{NP}$ ($\beta = 6.0, 6.2$).

\begin{figure}
\begin{center}
\epsfxsize=350pt\epsfbox{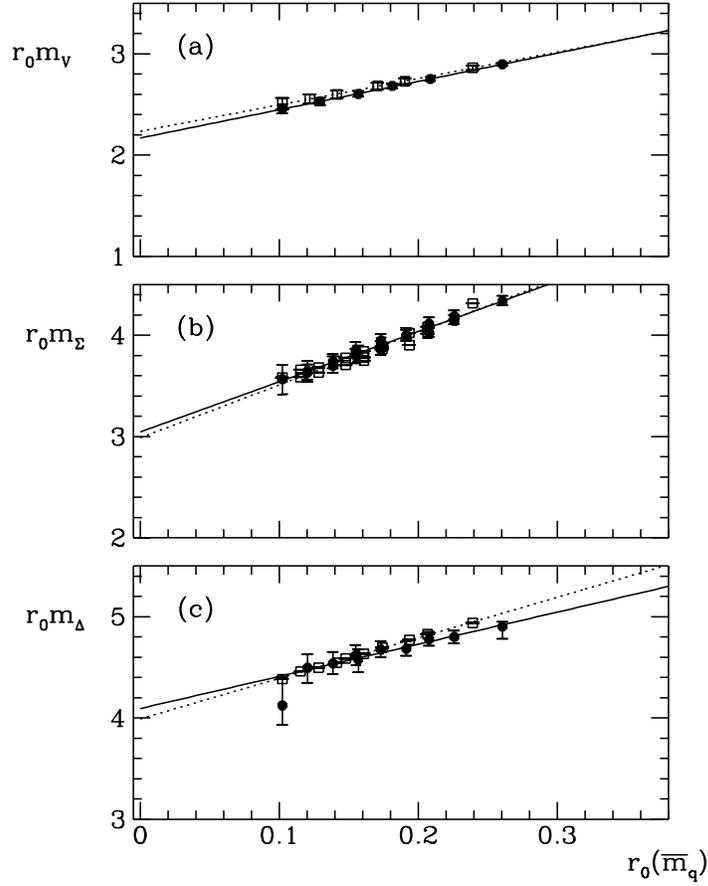}
\end{center}
\caption{ Data~\protect\cite{ukqcd2000} for (a): vector mesons, (b):
$\Sigma$-like baryons and (c): $\Delta$-like baryons is plotted
against the average value of the masses of the component quarks
$\bar{m_q} = (\tilde{m}_{q,1} + \tilde{m}_{q,2})/2$ (vector meson) and
$\bar{m_q} = (\tilde{m}_{q,1} + \tilde{m}_{q,2} + \tilde{m}_{q,3})/3$
(baryons).  Squares and circles denote the NP data at $\beta = 6.2$
and $\beta = 6.0$ respectively, and the lines correspond to the fits
of
Eqs.~(\protect\ref{eq:fit_chi_V})-(\protect\ref{eq:fit_chi_Delta})}.
\label{fig:chiral}
\end{figure}

Since the quark propagators were computed at values of the quark mass
in the region of the strange quark mass, it is necessary to extrapolate in
the quark mass to obtain results at the $u$- and $d$-quark masses, and
interpolate to obtain results at the $s$-quark mass.  The bare, or
unrenormalised quark mass, is related to the hopping parameter
$\kappa$ through
\begin{equation}
m_q = \frac{1}{2a} \left( \frac{1}{\kappa} - \frac{1}{\kappa_{\rm
crit}} \right).
\end{equation}
The bare quark mass must be rescaled in the ${\cal O}(a)$-improved
theory so that spectral quantities approach the continuum limit with
${\cal O}(a^2)$,\cite{luscher96}  
\begin{equation}
\tilde{m}_q = m_1 (1 + b_m am_q),
\end{equation}
where the perturbative one-loop value of $b_m$ is used.\cite{sint97}
To extrapolate the results to the physical quark masses, the following
ansatz is employed:
\begin{eqnarray}
m_{\rm PS}^2 & = & B(\tilde{m}_{q,1} +
\tilde{m}_{q,2}\label{eq:fit_chi_PS}\\ m_{\rm V} & = & A_{\rm V} +
C_{\rm V} (\tilde{m}_{q,1} + \tilde{m}_{q,2})\label{eq:fit_chi_V} \\
m_{\rm Oct} & = & A_{\rm Oct} + C_{\rm Oct} (\tilde{m}_{q,1} +
\tilde{m}_{q,2} + \tilde{m}_{q,3})\label{eq:fit_chi_N}\\ m_{\rm Dec} &
= & A_{\rm Dec} + C_{\rm Dec} (\tilde{m}_{q,1} + \tilde{m}_{q,2} +
\tilde{m}_{q,3}),\label{eq:fit_chi_Delta}
\end{eqnarray}
where the subscripts PS, V, Oct and Dec refer to the pseudoscalar
meson, vector meson, octet baryons ($\Sigma$- and $\Lambda$-like) and
decuplet baryons ($\Delta$-like) respectively.  The critical hopping
parameter corresponds to the value of $\kappa$ for which $m_{\rm PS}$
vanishes.  The quality of the chiral extrapolations of the
vector, $\Sigma$ and $\Lambda$  masses for the $\csw = \mbox{NP}$ data is
shown in Figure~\ref{fig:chiral}.

\begin{figure}
\begin{center}
\epsfxsize=300pt\epsfbox{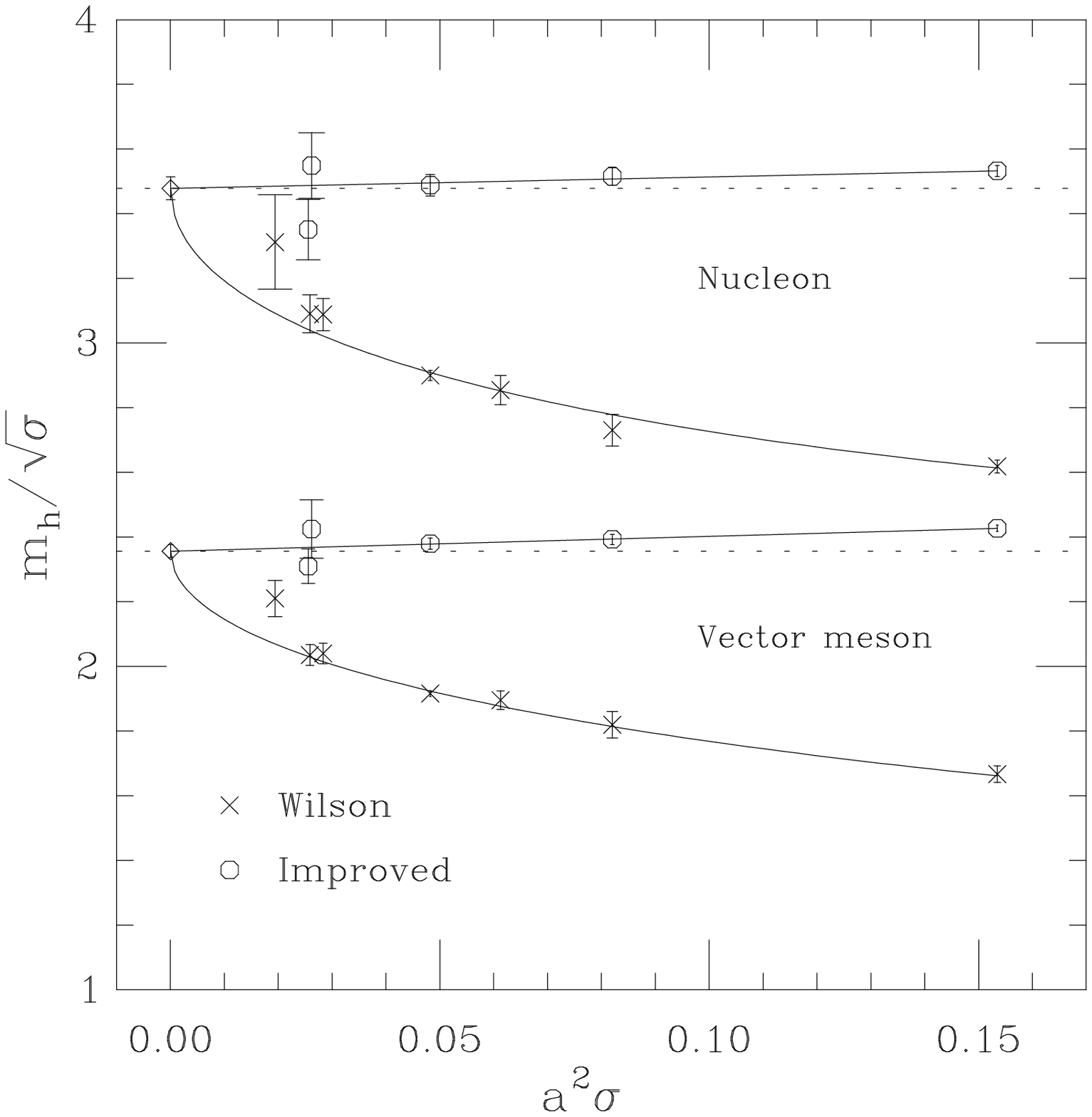}
\end{center}
\caption{The hadron spectrum using the Wilson and non-perturbatively
improved fermion actions is shown against $a^2$ at fixed $m_{\rm
PS}/m_{\rm V}$.\protect\cite{edwards97}  The scale is set from the
string tension.}
\label{fig:discrete}
\end{figure}

The $\csw = \mbox{TAD}$ data has in principle a remnant ${\cal O}(a)$
discretisation error, whilst the $\csw = \mbox{NP}$ data is fully
${\cal O}(a^2)$-improved.  In the continuum extrapolations, UKQCD
performs a simultaneous fit to both the NP and TAD data.
In order to investigate the approach to the continuum limit, it is not
necessary to study the chirally extrapolated values.  Indeed, a clear
demonstration of the efficacy of improvement can be seen by looking at
the lattice-spacing dependence of hadron masses at a fixed
$m_{\pi}/m_{\rho}$ ratio,\cite{edwards97} shown in
Figure~\ref{fig:discrete}.

\begin{figure}
\begin{center}
\epsfxsize=350pt\epsfbox{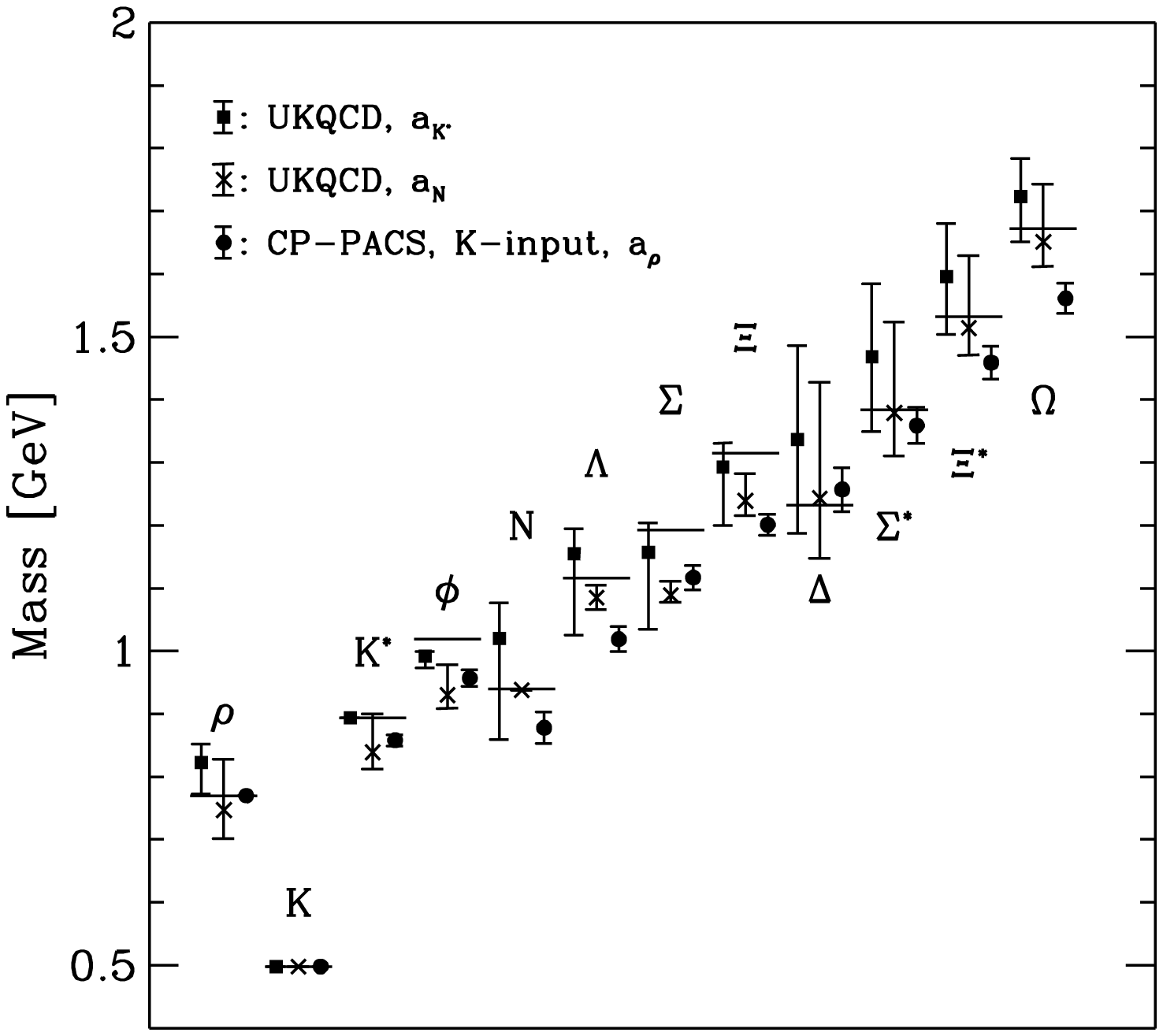}
\end{center}
\caption{The quenched light hadron spectrum computed in the ${\cal
O}(a)$-improved theory~\protect\cite{ukqcd2000} and the comparison to
results obtained using the unimproved Wilson action (full
circles).\protect\cite{cppacs2000} The levels of the experimental points are
denoted by solid lines}\label{fig:ukqcd_spectrum}
\end{figure}
The final UKQCD results for the quenched light-hadron spectrum,
together with the CP-PACS results using the standard Wilson fermion
action, are shown in Figure~\ref{fig:ukqcd_spectrum}.  The different
UKQCD plotting symbols correspond to determining the lattice spacing
by requiring that either $K^*$ or $N$ have its physical value.  Both
calculations support the assertion that the quenched light-hadron
spectrum agrees with experiment at the 10\% level.  Now that we have
established the reliability with which we can compute the known hadron
masses, it is time to investigate the predictive power of lattice
QCD.  We begin by considering the glueball spectrum.

\subsection{Glueball Spectrum}
The gluon self-coupling that distinguishes QCD from the Abelian QED
admits the existence of purely gluonic bound states, or glueballs.
Indeed, glueballs are the only true states in quenched QCD!  The good
agreement of the quenched light-hadron spectrum with the experimental
value was perhaps not surprising; Zweig's rule tells us that hadronic
decays involving the annihilation of the initial quarks are highly
suppressed.  Zweig's rule also suggests that glueball mixing with
quark states should be similarly suppressed.  Thus a quenched
calculation of the glueball spectrum is very important, and can yield
crucial information to guide experiment.

The calculation of the glueball spectrum has been plagued by two
inter-related problems.  Firstly, the glueballs are relatively heavy
and thus the correlation functions die rapidly at increasing temporal
separations.  Secondly the glueball correlators are subject to large
fluctuations independent of separation.  In consequence, the
signal-to-noise ratio for glueball correlators is very poor.

\begin{figure}
\begin{center}
\epsfxsize=200pt\epsfbox{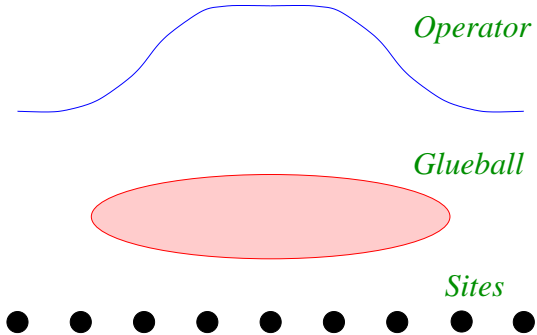}
\end{center}
\caption{Construction of improved glueball operators, with the aim of
increasing the overlap with the ground state.}\label{fig:fuzzing_idea}
\end{figure}

Calculations of the spectrum using the standard Wilson gluon action,
Eq.~(\ref{eq:gauge_action}), have emphasised the construction of
improved gluonic operators which more correctly describe the
ground-state glueball wave function, as illustrated in
Figure~\ref{fig:fuzzing_idea}.
\begin{figure}
\begin{center}
\epsfxsize=350pt\epsfbox{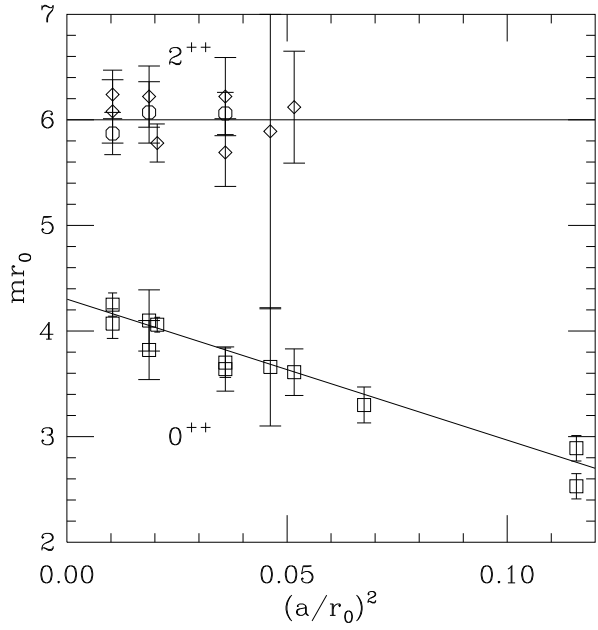}
\end{center}
\caption{The continuum extrapolation of the masses of the $J^{PC} =
0^{++}$ and $2^{++}$ states.\protect\cite{michael97} The different
plotting symbols for the $2^{++}$ states correspond to the lattice
operators $T_2$ (octagons) and $E$ (diamonds).  The lines represent
linear continuum extrapolations in $a^2$.}\label{fig:glueball_states}
\end{figure}
In the case of the determination of the spectrum of states at rest,
the rotation group used in the construction of the continuum glueball
operators is reduced to the cubic group of the lattice.  The different
components of, for example, the $J^{PC} = 2^{++}$ glueball lie in the
$E^{++}$ and $T_2^{++}$ representations of the cubic group.  As the
continuum limit is approached, the masses obtained from the different
representations of the cubic group should become degenerate,
signifying the restoration of rotation symmetry.  Despite the
discretisation errors of the standard Wilson gauge action being ${\cal
O}(a^2)$, the lightest glueball state is subject to much larger ${\cal
O}(a^2)$ discretisation errors than the spectrum for quark states, as
shown in Figure~\ref{fig:glueball_states}.\cite{michael97}

\begin{figure}
\begin{center}
\epsfxsize=250pt\epsfbox{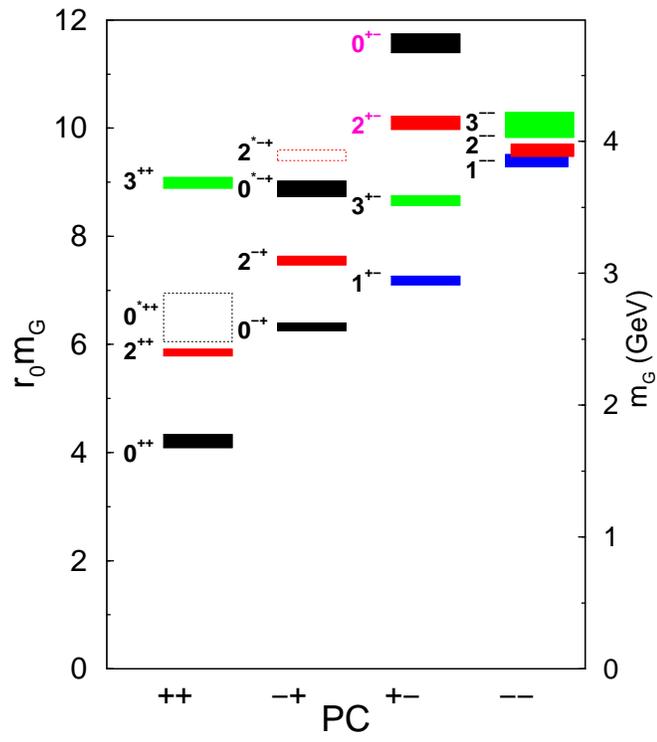}
\end{center}
\caption{The quenched glueball spectrum is shown after extrapolation
to the continuum limit.\protect\cite{morningstar99}}\label{fig:cont_glueballs}
\end{figure}
To improve upon these calculations, Morningstar and
Peardon~\cite{morningstar95} chose to employ an ${\cal
O}(a^2)$-improved gluon action, having discretisation errors of ${\cal
O}(a^4)$;note that it is only necessary to consider operators of even
dimension in the construction of the gluonic part of the action.
Unfortunately, the relatively large mass of the glueball states in
lattice units allows only a couple of time slices to be used in
extracting the glueball masses.  They then observed that one could
employ a relatively coarse lattice in the spatial directions to
produce a reasonable approximation to the glueball wave function,
whilst employing a finer lattice in the temporal direction to enable
the isolation of the ground state and excited state masses in each
channel.\cite{morningstar96}

The anisotropic lattice is implemented through the choice of different
coupling constants for the space-space and space-time plaquettes in
the gluonic action Eq.~\ref{eq:gauge_action}; this involves a
non-trivial tuning, since the bare couplings are renormalised.  The
resulting glueball spectrum,\cite{morningstar99} after extrapolation to
the continuum limit, is shown in Figure~\ref{fig:cont_glueballs}.
\subsection{Exotic Hadrons}
The search for hadrons with excited glue is one of the primary goals
of the $N^*$ programme at CEBAF.  There has been a flurry of recent
activity looking for exotic states, and in particular exotic mesons,
in the lattice community; lattice gauge theory aficionados generally
try to comprehensively understand the mesonic sector before venturing
into the realm of baryons.

\subsubsection{Exotics and Hybrids}
Within the quark model, the charge conjugation $C$ and parity $P$ of a
meson are related to the spin $S$ and orbital angular momentum $L$ through
\[
P = (-1)^{L+1} \,; C = (-1)^{L+S}.
\]
States not conforming to these relations are called exotics, and
examples are the states with
\[
J^{PC} = 1^{-+}, 0^{+-}, 2^{+-}.
\]
An exotic can be formed in two ways.  Firstly, as a
quark-antiquark-glue state, which we call a hybrid.  Secondly as a
bound state of two quarks and two antiquarks.  In this section I will
discuss lattice studies of hybrid states.  These studies have been
given increased impetus by experimental observations of $1^{-+}$
resonance states in the region of
$1.4\,\mbox{GeV}$.\cite{E852:97,VES93}

\subsubsection{Hybrid interpolating operators}
The usual interpolating operator for the pion is
\begin{equation}
{\cal O}_{\pi}(\vec{x}, t) = \bar{\psi}(\vec{x},t) \gamma_5
\psi(\vec{x},t).
\end{equation}
In the case of the hybrid state $1^{-+}$, a possible interpolating
operator would be
\begin{equation}
{\cal O}_{1^{-+}}(\vec{x},t) = \bar{\psi}(\vec{x},t) \gamma_i
F_{ij}(\vec{x},t) \psi(\vec{x},t)
\end{equation}
where $F_{ij}$ is a lattice discretisation of the field-strength
tensor constructed in Figure~\ref{fig:clover}.  In practice to get any
sort of signal for hybrid mesons, it is necessary to use interpolating
operators in which the quark and antiquark are separated in space.
The signals are inevitably much noisier than for the pseudoscalar and
vector meson states.  To see why, we note that the hybrid correlator
$C_{\rm hyb}(t)$ decays exponentially with the mass of the hybrid,
\begin{equation}
C_{\rm hyb}(t) \simeq e^{-m_{\rm hyb} t}.
\end{equation}
In contrast, the correlator for the variance is that of the square of
the interpolating operator
\begin{equation}
C_{\sigma^2}(t) = \sum_{\vec{x}} \langle |{\cal O}_{\rm
hyb}(\vec{x},t)|^2 
|{\cal O}_{\rm hyb}(\vec{0},0)|^2 \rangle.
\end{equation}
Typically, $|{\cal O}_{\rm hyb}|^2$ is an interpolating operator for two
pions, and therefore the signal-to-noise ratio with increasing
temporal separations increases as
\begin{equation}
\frac{\rm signal}{\rm noise} \simeq e^{-(m_{\rm hyb} - m_{\pi})t}.
\end{equation}
Since the masses of the hybrids are relatively large, the signal
quickly is lost in the noise.  In the case of the glueball
calculations, the situation is particularly severe since the square of
the glueball operator has a non-zero vacuum expectation value, leading
to the constant noise alluded to earlier.

There have been recent calculations of the light-quark hybrid
spectrum, and in particular of the $1^{-+}$ state, in both the quenched
approximation,\cite{lacock96,milc98} and in full
QCD.\cite{milc97,lacock98}  These calculations all find a
$1^{-+}$ mass around $2~{\rm GeV}$, far larger than that of the
experimental candidates.  A possible resolution is that this resonance is
actually a four-quark state; we will return to this interpretation in
Section~\ref{sec:multihadron}.

\subsection{The $N^*$ Spectrum}
\begin{figure}
\begin{center}
\epsfxsize=250pt\epsfbox{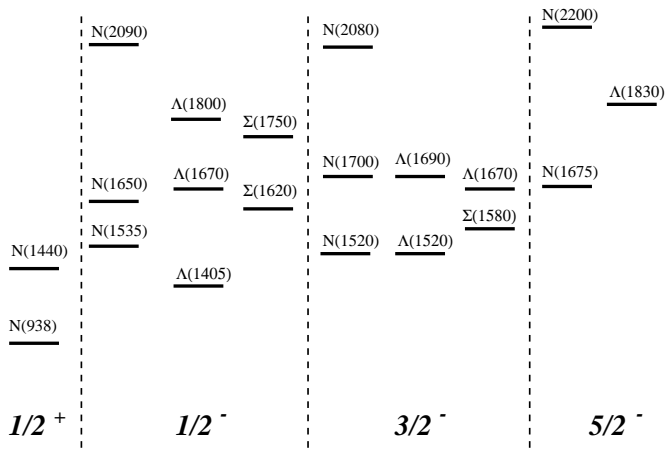}
\end{center}
\caption{Schematic showing the observed $N^*$ spectrum, labelled by
$J^P$.}\label{fig:observed_nstar}
\end{figure}

The measurement of the excited nucleon spectrum reveals the full
$SU(3)$ nature of QCD, and is a critical part of the experimental
programme at CEBAF. The observed $N^*$ spectrum is shown schematically
in Figure~\ref{fig:observed_nstar}. 
The nucleon $N(938)$ has positive parity; its parity partner, with
negative parity, is the $N(1535)$.  The usual nucleon interpolating
operators employed in lattice calculations are 
\begin{equation}
N_{\alpha}  =  \epsilon^{ijk} (u^i C \gamma_5 d^j) u^k_{\alpha}.
\end{equation}
On forming the time-sliced correlator, both positive and negative
parity states can contribute to the correlation function.  However, we
can perform a parity projection
\begin{equation}
C(t) = \sum_{\vec{x}} \langle 0 \mid N_{\alpha}(\vec{x},t) (1 \pm
\gamma_0)_{\alpha\beta} \bar{N}_{\beta}(0) \mid 0 \rangle
\end{equation}
to project out the forward-propagating positive (+ sign) or negative
(- sign) parity states, with states of the opposite parity propagating
in the negative direction.  On a periodic lattice, our correlator
contains both the forward- and backward-propagating states, and we
rely on a sufficiently long temporal extent to our lattice to
delineate the two parities.  

Recently, there have been two lattice calculation of this mass
splitting.  The first~\cite{lee99} employs a highly improved fermion
action, the $D_{\chi}34$ action of Hamber and Wu.\cite{hamber84} Not
only do the authors extract the mass of the $J=1/2^-$ state, but also
find a signal for the $J = 3/2^-$ state.  The second
calculation~\cite{riken00} employs domain-wall fermions; here the
authors argue that, since the $J = 1/2^+$ and $J = 1/2^-$ state are
degenerate in an unbroken chirally symmetric theory, the use of an
action having a possessing exact chiral symmetry, even at non-zero
lattice spacing, is crucial in correctly extracting the splitting.

\begin{figure}
\begin{center}
\epsfxsize=250pt\epsfbox{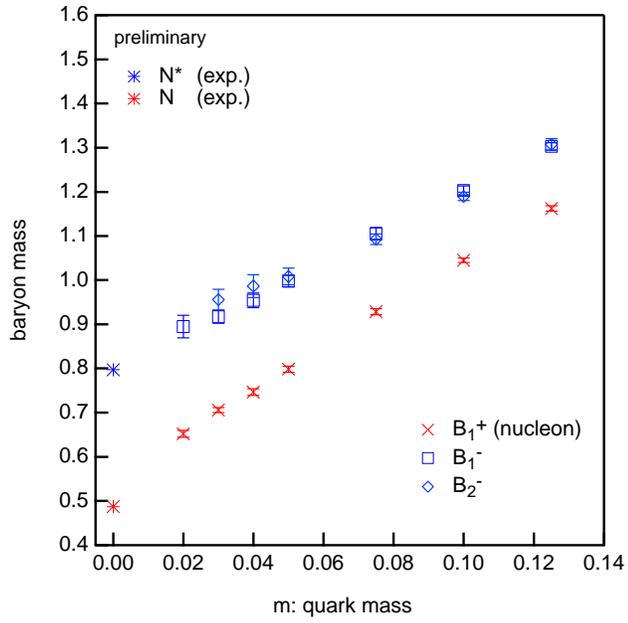}
\end{center}
\caption{The masses of the nucleon (N) and its negative-parity partner
in lattice units ($a^{-1} \simeq 1.9~\mbox{GeV}$ using the $\rho$ mass
to set the scale).\protect\cite{riken00}}\label{fig:nstar}
\end{figure}
Both calculations find mass splittings between the positive- and
negative-parity states in accord with experiment, though with still
substantial systematic and statistical uncertainties.  The
masses of the $J = 1/2^+$ (nucleon) and $J = 1/2^-$ states using DWF
at a fixed value of the lattice spacing are shown in
Figure~\ref{fig:nstar}.

\section{Hadron Structure}
As well as enabling the calculation of the hadron spectrum, lattice
QCD enables the study of the distribution of the quarks and gluons
within hadrons.  Information about these is contained in the form
factors, and in the quark and gluon structure functions.  These
calculations have in common the determination of some (local) hadronic
matrix element, so we will begin this section with a discussion of the
lattice technology of determining matrix elements.

\subsection{Hadronic Matrix Elements}
The paradigm calculation is that of $f_{\pi}$, the pion decay constant
defined through
\begin{equation}
\langle 0 \mid A_{\mu} \mid \pi \rangle  = i p_{\mu} f_{\pi},
\end{equation}
where $A_{\mu} \equiv \bar{\psi} \gamma_{\mu} \gamma_5 \psi$ is the
axial vector current.  This matrix element we can obtain as a
by-product of the determination of $m_{\pi}$, and our analysis will
follow that of Section~\ref{sec:spectrum_recipe}.  We construct the
correlator
\begin{equation}
C(t) = \sum_{\vec{x}} \langle 0 \mid A^{\rm lat}_4(\vec{x},t) A^{\rm
lat}_4(0) \mid 0.  \rangle,
\end{equation}
where $A^{\rm lat}$ is a lattice discretisation of the continuum
axial-vector current.
Inserting a complete set of states between the two interpolating
operators, we obtain
\begin{eqnarray}
C(t) & = & \sum_{\vec{x}} \sum_P \frac{1}{(2 \pi)^3} \int \frac{d^3\,
\vec{p}}{2 E(\vec{p})} \langle 0 | A^{\rm lat}_4(\vec{x}, t) |
P(\vec{p}) \rangle \langle P(\vec{p}) | {A^{\rm lat}_4}^{\dagger}(0) |
0 \rangle \nonumber\\ 
& = & \sum_{\vec{x}} \sum_P \frac{1}{(2 \pi)^3}\int\frac{d^3 \,
\vec{p}}
{2 E(\vec{p})} e^{-i E t + i
\vec{p}\cdot \vec{x}} \langle 0| A^{\rm lat}_4(0) | P(\vec{p}) \rangle
\langle P(\vec{p}) | {A^{\rm lat}_4}^{\dagger}(0) | 0 \rangle\nonumber\\
& = & \sum_P \int \frac{d^3 \, \vec{p}}{2 E(\vec{p})} e^{-iEt} \,
\delta^{(3)} (\vec{p}) \, \langle 0| A^{\rm lat}_4 | P(\vec{p})
\rangle \langle P(\vec{p}) | {A^{\rm lat}_4}^{\dagger} | 0 \rangle
\nonumber\\ & \stackrel{t \rightarrow \infty}{\longrightarrow}&
\frac{1}{2} \, m_{\pi} {f^{\rm lat}_{\pi}}^2 \, e^{- \, m_{\pi} \, t}
\, + \mbox{\textit{excited states}}
\end{eqnarray}
The lattice discretisation has provided both a cut-off, and a
renormalisation scheme.  We, of course, want $f_{\pi}$ is some
familiar continuum scheme, such as $\overline{\rm MS}$.  To provide us
with that, we need to compute the matching coefficient $Z_A$ for the
axial vector current, such that
\begin{equation}
A^R_{\mu} = Z_A A^{\rm lat}.
\end{equation}
The determination of the matching coefficient $Z$ is one of the most
delicate, and generally onerous, tasks of any calculation of hadronic
matrix elements.  The lattice formulation reproduces continuum QCD as
the lattice spacing approaches zero, and thus the anomalous dimensions
of the operators in the lattice formulation matches those of the
continuum operators.  However, the replacement of the continuum
Lorentz symmetry by the hypercubic symmetry of the lattice, and the
lack of chiral symmetry in most fermion formulations, complicates the
calculation of the matching coefficients enormously, even for such a
simple operator as the axial vector current.  In particular, the
lattice allows mixing with higher dimension operators, combined with
appropriate powers of the lattice spacing $a$.  An important element
of the improvement programme is finding a combination of lattice
operators such that matrix elements are free of ${\cal O}(a)$, or
higher, discretisation errors.

In principle, we can compute $Z$ in perturbation theory.  Perturbation
theory using the bare lattice coupling $g$ as an expansion parameter
apparently fails,  resulting in very large perturbative corrections.
The bulk of these large corrections can be identified as lattice
artifacts, the ``tadpole'' contributions arising from the ${\cal
O}(g^2)$ term in the expansion of the link variable,
Eq.~\ref{eq:link_variable}.  These terms can effectively be resummed
through the expansion in terms of a renormalised  coupling
constant.\cite{lm93} This prescription is precisely that used in the
specification of the tadpole-improved clover coefficient of
Eq.~\ref{eq:csw_tad}.

An alternative route is to attempt to determine the matching
coefficients, and improvement coefficients, non-perturbatively through
the imposition of chiral Ward identities.\cite{alpha96} Indeed, this
prescription enables, in principle, the elimination of all ${\cal
O}(a)$ discretisation effects from on-shell quantities, providing
appropriate renormalisation conditions can be found.

Both these routes require considerable effort, and uncertainty in the
calculation of matching and improvement coefficients is a major
uncertainty in the calculation of hadronic matrix elements.  Let me
conclude this subsection by noting that the chiral fermion actions of
Section~\ref{sec:chiral} are automatically ${\cal O}(a)$-improved, and
admit a smaller degree of operator mixing.  This may be prove a
substantial advantage for these formulations.

\subsection{Nucleon Form Factors}
The electric and magnetic form factors of the nucleon are among the simplest
quantities that contain information about the structure of the
nucleon, and are measured in electron proton scattering.  They are
related to the matrix elements of the vector current $J_{\mu}$ through
\begin{equation}
\langle p', s' \mid J_{\mu}(q) \mid p,s\rangle = \bar{u}(p',s')
\left[ \gamma_{\mu} F_1(q^2) + i \sigma_{\mu\nu} \frac{q^{\nu}}{2
m_N} F_2(q^2) \right] u_p(p,s),
\end{equation}
where $q = p - p'$ is the momentum of the photon probe.  Note that
$F_1$ and $F_2$ satisfy
\[
F_1(0) = 1;\quad F_2(0) = \mu - 1
\]
where the former result expresses current conservation, and
$\mu$ is the magnetic moment of the nucleon.  For point-like
particles, both quantities would be constant, and therefore they are a
measure of the spatial extent of the nucleon.  Rather than quoting
these quantities directly, it is usual to form the Sach's form factors
\begin{eqnarray}
G_E(q^2) & = & F_1(q^2) + \frac{q^2}{4 m_N^2} F_2(q^2)\\
G_M(q^2) & = & F_1(q^2) + F_2(q^2),
\end{eqnarray}
where we note that $q^2$ is space-like.  

\begin{figure}
\begin{center}
\epsfxsize=250pt\epsfbox{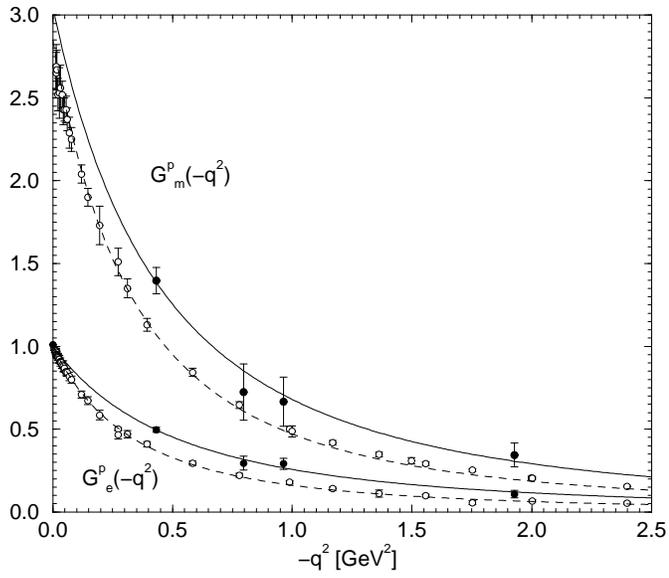}
\end{center}
\label{fig:nucleon_form_factor}
\caption{The solid points are lattice determinations of the electric
and magnetic form factors of the proton in the quenched approximation
to QCD.\protect\cite{capitani98} The open circles are the experimental
measurements.\protect\cite{cristov96} The lines represent the dipole
fits of Eq.~\protect\ref{eq:ff_dipole} to the experimental and lattice
data.}
\end{figure}
Phenomenologically, it is usual to parameterise the form factors
through the vector dominance model by a dipole fit
\begin{eqnarray}
G^p_E(q^2) & \sim & G^p_M(q^2)/\mu^2 \sim G^N_M(q^2) \sim
\frac{1}{\left( 1 - q^2/m_V^2 \right)^2}\nonumber\\
G^n_E(q^2) & \sim & 0.\label{eq:ff_dipole}
\end{eqnarray}
A recent lattice determination of the proton electromagnetic form
factors~\cite{capitani98} is shown in
Figure~\ref{fig:nucleon_form_factor}, together with the experimental
data.\cite{cristov96} The lines are fits to the data using the dipole
forms of Eq.~\ref{eq:ff_dipole}.

\subsection{Hadronic Structure Functions}
The hadronic structure functions, describing the distribution of
quarks and gluons inside, say, a nucleon in inclusive processes, are
related to the hadronic tensor
\begin{equation}
W_{\mu\nu} = \frac{1}{4\pi} \int d^4x e^{iq \cdot x} \langle
N(p, s) |
J_{\mu}(x) J_{\nu}(0) | N(p,s) \rangle, \label{eq:hadronic_tensor}
\end{equation}
where $J_{\mu}$ is the electroweak current, and $p, s$ are the
nucleon momentum and spin respectively.  Decomposing $W_{\mu\nu}$
according to the possible Lorentz structures yields four structure
functions, two spin-averaged, $F_{1,2}(x,Q^2)$ and two
spin-dependent, $g_{1,2}(x,Q^2)$, where $x$ is the Bjorken variable,
and $Q^2 = - q^2$. Phenomenologically, these are determined in Deep
Inelastic Scattering, and parameterised at some reference energy
scale.

Near the light cone, $x^2 \sim 0$, the structure functions can be
expanded using the operator product expansion (OPE) in terms of the
matrix elements of certain local operators, of twist (dimension -
spin) two, together with Wilson coefficients calculable in
perturbation theory.  Historically, it is these matrix elements that
are computed in lattice simulations, since the non-zero lattice
spacing in our calculations precludes the measurement of the currents
in Eq.~\ref{eq:hadronic_tensor} at sufficiently small separations, and
more fundamentally it is unclear how to extract this quantity at
light-like separations in Euclidean space.

The local matrix elements are related to the $x$-moments of the structure
functions.  In principle, we can recover the full $x$-dependence of the
structure functions by measuring increasing moments.  The simplest
operators are the non-singlet operators for both the unpolarised and
polarised structure functions,
\begin{eqnarray}
{\cal O}_{\mu_0\dots\mu_n} & = & \bar{\psi} \gamma_{\mu_0} i
D_{\mu_1}\dots i D_{\mu_n} \tau \psi\nonumber\\
{\cal O}^5_{\mu_0\dots\mu_n} & = & \bar{\psi} \gamma_5 \gamma_{\mu_0}
i D_{\mu_1}\dots i D_{\mu_n} \tau \psi,
\end{eqnarray}
where symmetrisation of indices and removal of traces is understood,
and the $\tau$ is an SU(2) flavour matrix.  The first few moments
moments for the nucleon have been measured by several
groups.\cite{beccarini95,gockeler95,brower96,dolgov98}

\begin{figure}
\begin{center}
\epsfxsize=250pt\epsfbox{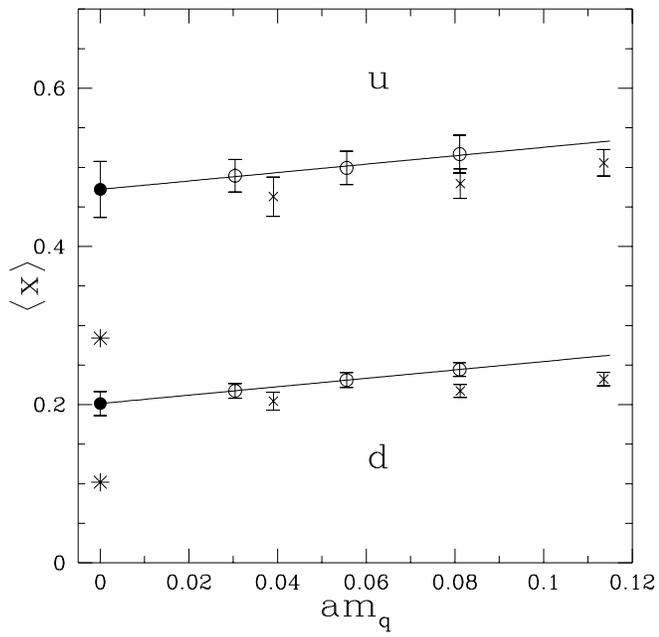}
\end{center}
\caption{The first moment of the unpolarised $u$ and $d$ quark
distributions in the nucleon as a function of the quark mass at $\beta
= 6.0$.\protect\cite{gockeler97} The circles and crosses denote results with
the Wilson and clover fermion actions
respectively.  The bursts denote the CTEQ
results.\protect\cite{cteq95}}\label{fig:str_unpol}
\end{figure}

The lowest moment of the unpolarised quark distribution has a
particularly simple interpretation in terms of the momentum fraction
carried by the quarks.  Figure~\ref{fig:str_unpol}
shows the first moment of the $u$ and $d$ quark distributions in the
quenched approximation;\cite{gockeler97} both distributions are
somewhat higher than phenomenological expectations.  Unfortunately,
it is unclear the extent to which the calculation of the first few
moments of the structure functions enables a useful picture of the
$x$-dependence of the structure functions, and a means of performing a
direct computation of the hadronic tensor,
Eq.~\ref{eq:hadronic_tensor}, would be invaluable.

The study of the flavour-singlet sector involves consideration of both
the quark and gluon distributions, which mix.  Computationally, these
studies are much more demanding, but of tremendous interest since they
venture beyond the simple valence picture of the nucleon.  More
recently, there have been attempts to investigate the r\^{o}le of
higher-twist contributions,\cite{capitani99} to which upcoming
experiments at JLAB should be particularly sensitive.

\section{The Nucleon-Nucleon Interaction}\label{sec:multihadron}
In the preceding, we have used lattice gauge theory to acquire a fundamental
understanding of the internal structure of an isolated hadron from
first principles.  It is natural to aim at a similar understanding of
the interactions between hadrons, and in particular of the
nucleon-nucleon interaction, the very foundation of nuclear physics.

Understanding the strong interaction in multi-hadron systems from
lattice QCD is a notoriously difficult problem.  Multi-hadron states
involve the computation of a four-point function and are relatively
massive, and therefore the corresponding correlation functions quickly
vanish into noise at increasing temporal separations.  Furthermore,
multi-hadron systems are large, and therefore the spatial extent of
the lattice needs to be correspondingly larger than that used in
hadronic spectroscopy.  Finally, the use of a Euclidean lattice
obscures the the extraction of the phase information of the full
scattering matrix.\cite{maiani90} Despite these difficulties, the
problem is fundamental and compelling.

Historically, there have been two approaches to this problem within
lattice QCD.  The first aimed to extract certain quantitative
parameters of the hadron-hadron interaction by direct lattice
simulation.  L\"{u}scher\cite{Lue86,Lue91a,Lue91b} exploited the
finite-size dependence to extract a discrete set of $s$-wave
scattering lengths.  The was thoroughly tested within an
$O(4)$-symmetric $\phi^4$ model~\cite{Goe94}, and scattering lengths
have been computed within QCD for pions,\cite{Fuk95,Fuk99} and for
nucleons.\cite{Fuk95}  Fiebig \textit{et al.}~\cite{Fie00a}
explored the $I = 2$ $\pi-\pi$ system by extracting a residual
interaction potential, and then proceeded to compute the scattering
phase shifts which were compared with experiment.

\begin{figure}
\begin{center}
\epsfxsize=250pt\epsfbox{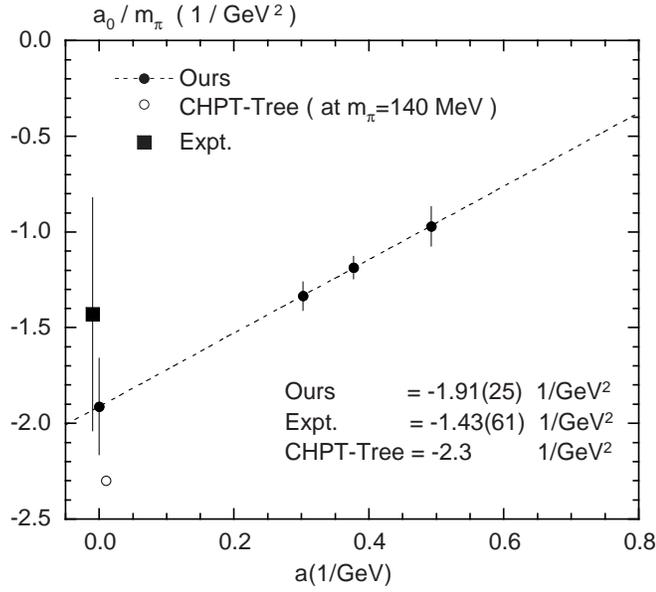}
\end{center}
\caption{The $I=2$ pion scattering length obtained in the
quenched approximation to QCD is shown as a function of the lattice
spacing, and in the continuum limit.\protect\cite{Fuk99}  Also shown
as the open circle is the current-algebra prediction for $m_{\pi} =
140\,\mbox{MeV}$.}\label{fig:swave_pion}
\end{figure}
The extraction of the $s$-wave scattering lengths of the $\pi-\pi$
interaction has been encouraging, as illustrated in
Figure~\ref{fig:swave_pion} where the $s$-wave scattering length for
the isospin $I=2$ pion-pion interaction is shown.\cite{Fuk99} The
investigation of the $N-N$ system is more problematical.  Here the
scattering lengths are of the order of 10~fm, rather than less than
1~fm as is the case for the $\pi-\pi$ interaction.  Therefore, while
lattice calculations do indeed find scattering lengths for the $N-N$
interaction considerably larger than those for the $\pi-\pi$ and
$\pi-N$ interaction,\cite{Fuk95} this approach is limited by our
present inability to simulate on lattice sizes of the order of $10 -
20~\mbox{fm}$, and at physical values of the pseudoscalar mass.
\begin{figure}
\begin{center}
\epsfxsize=250pt\epsfbox{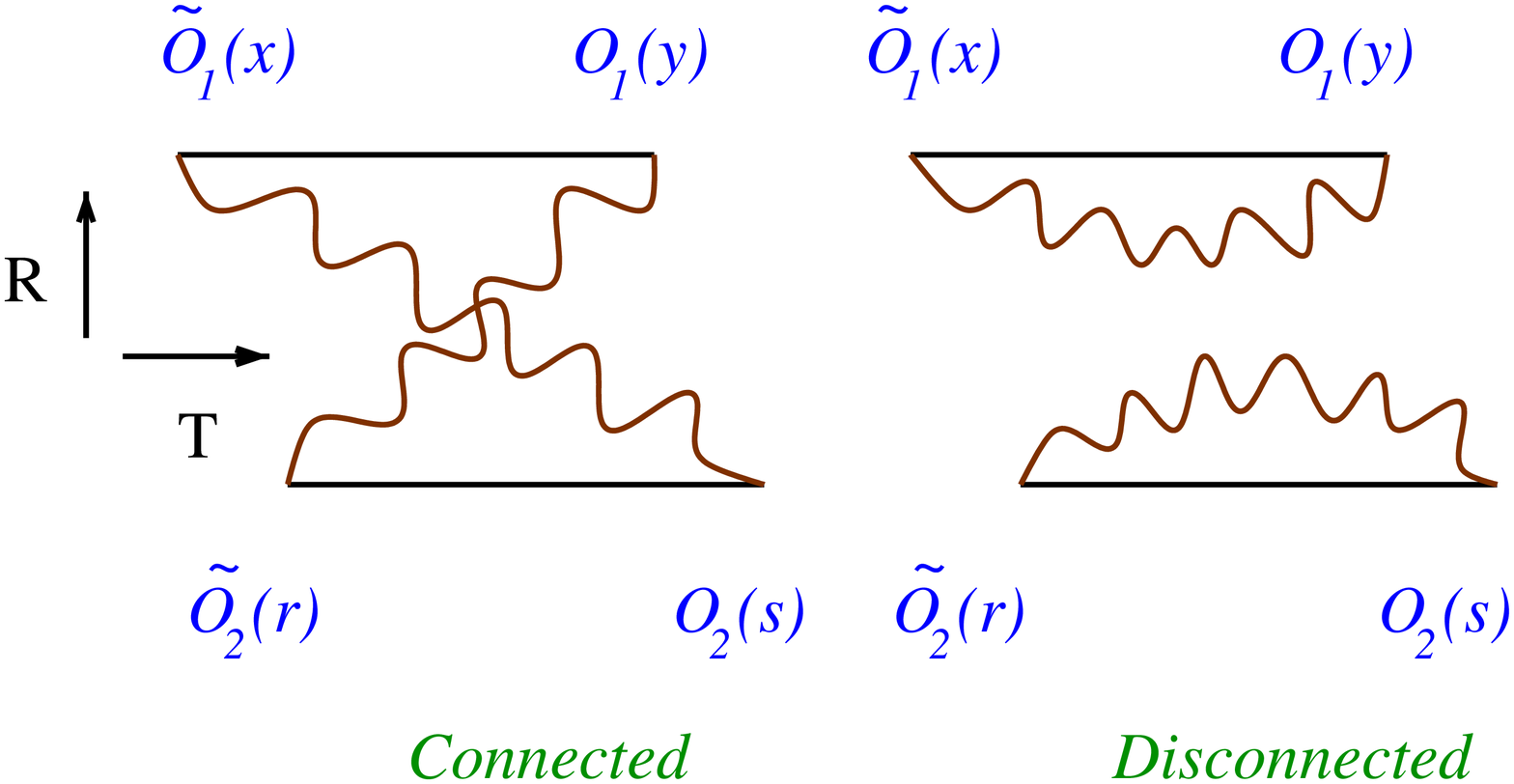}
\end{center}
\caption{The connected (flavour-exchange) and disconnected contributions to the
interaction between two heavy-light mesons are shown.  The solid lines
denote the heavy quarks, whilst the curly lines denote the propagation
of the light-quarks.}\label{fig:fourpoint}
\end{figure}

The second approach is motivated by the realisation that important
insight into the nucleon-nucleon interaction can be gleaned by
studying the interactions of a much simpler system, that of two
heavy-light mesons, with static heavy quarks.\cite{Ric89,Ric90} Such a
system exhibits most of the salient features of the nucleon-nucleon
system, namely quark exchange, flavour exchange and colour polarization.
The interaction between the mesons can be understood by the study of
the four-point functions, shown in Figure~\ref{fig:fourpoint}:
\begin{equation}
C^{(4)}(x,r;y,s) = \langle O_1(y) O_2(s) \tilde{O}_1^{\dagger}(x)
\tilde{O}_2^{\dagger}(r) \rangle.\label{eq:fourpoint}
\end{equation}
Here each operator can either be that for a pseudoscalar ($P$) or for
a vector ($V$).  The range of the interaction can be assessed by
forming the $z$- and $t$-sliced sum,
\begin{equation}
\tilde{C}(R,T) = \sum_{t_1, t_2, \vec{a}_{\perp}}
C(x,r;y,s),\label{eq:fourpoint_range}
\end{equation}
where
\begin{eqnarray*}
x & = & (\vec{0}, t_1), \, y = (\vec{0}, t_1 + T)\\
r & = & (\vec{a}_{\perp}, R, t_2), \, s = (\vec{a}_{\perp}, R, t_2 +
T).
\end{eqnarray*}
At large distances $R$ and times $T$, the correlation function  is
dominated by the lightest state $\mid n \rangle$, of mass $M_n$, that
can be exchanged between the mesons, with
\begin{equation}
\tilde{C}(R,T) \sim e^{- M_n R} \langle 0 |
O_1(\vec{0},T) \tilde{O}_1^{\dagger}(\vec{0},0) | n\rangle
\times \langle n |O_2(\vec{0},T)
\tilde{O}_2^{\dagger}(\vec{0},0) | 0 \rangle.
\end{equation}
Can these correlation functions be constructed from the elemental
quark propagators of Eq.~\ref{eq:quark_prop}?  The connected diagram
of Figure~\ref{fig:fourpoint} requires the evaluation of the
propagator from one point on every time slice to every point on the
lattice; this is manageable.  Unfortunately, the computation of the
disconnected diagram requires the evaluation of all-to-all
propagators.  Nonetheless, the study of the connected diagram alone is
valuable.  It describes the flavour-exchange interaction, and lattice
simulations have shown that the interaction is indeed mediated by
meson exchange at large distances, and furthermore that the quantum
numbers of the exchanged particle are in accord with naive
expectations; for the process $PP \rightarrow PP$ a vector
meson is exchanged, whilst for the process $PV \rightarrow VP$ a
pseudoscalar particle is exchanged.\cite{Ric90} 

\begin{figure}
\begin{center}
\vspace{-1in}
\epsfxsize=350pt\epsfbox{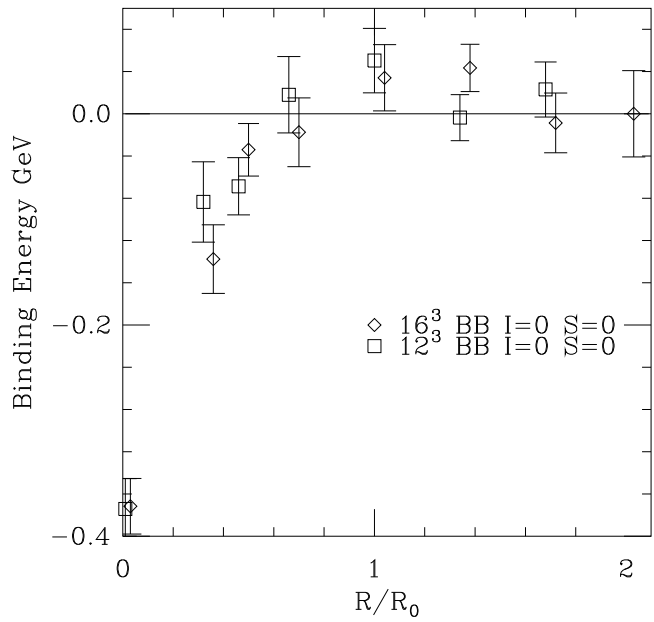}
\caption{Adiabatic potential between two heavy-light mesons, with the
light quarks in an isospin $I = 0$ and spin $S = 0$ configuration,
from reference~\protect\cite{Mic99}.  Measurements are obtained in the
quenched approximation to QCD, on lattices at $\beta = 5.7$.  The
separations are measured in units of $R_0 = 0.53~\mbox{fm}$, and the
different plotting symbols correspond to different lattice sizes.
}\label{fig:michael_binding}
\end{center}
\end{figure}
The especially attractive feature of heavy-light systems is that the
heavy quarks admit the definition of a relative coordinate, and
thereby a local adiabatic potential.  Recently, exploratory studies
have been made of this potential,\cite{Mih97,Mic99} and evidence for
nuclear binding sought, as illustrated in
Figure~\ref{fig:michael_binding}.  The investigation of this potential
is important and more feasible than the measurements of the scattering
lengths, because the large scattering lengths for the $N-N$ system
come from a short-range potential.  Furthermore, by exploring the
potential, we can discover the relative importance of gluon and meson
exchange contributions at various distances.  Understanding the nature
of this potential is also crucial to spectroscopy; are exotic mesons
predominantly quark-antiquark-glue hybrid states, or four-quark states?  

\section{Conclusions}
In these lectures I have tried to convey the power of lattice gauge
calculations as a means of understanding hadronic physics.  There are
very many areas that I have not addressed - topology and the role of
instantons, finite-temperature and finite-density phase transitions,
and, reaching beyond hadronic physics, the calculation of
weak-interaction matrix elements and even supersymmetry and gravity.
I trust that I have convinced you that lattice gauge theory provides
not only an \textit{ab initio} tool for obtaining quantitative results
(the spectrum, form factors etc.), but also provides a means of
increasing our conceptual understanding of the strong interaction, for
example through the study of the nuclear-nuclear force.

\section*{Acknowledgements}
I am grateful for helpful discussions with Stefano Capitani, Robert
Edwards, Rudolf Fiebig, Nathan Isgur, Xiandong Ji, Frank Lee, and
Stephen Wallace.  I also thank Jose Goity and staff for their
excellent HUGS workshop.  This work was supported by DOE contract
DE-AC05-84ER40150 under which the Southeastern Universities Research
Association (SURA) operates the Thomas Jefferson National Accelerator
Facility.


\section*{References}
\begin{thebibliography}{99}
\bibitem{wilson74} K.~Wilson, Phys.\ Rev.\ {\bf D10} (1974) 2445.
\bibitem{hmc} S.~Duane, A.D.~Kennedy, B.J.~Pendleton and D.~Roweth,
Phys.\ Lett.\ {\bf B195} (1987) 216.
\bibitem{nielsen81} H.B.~Nielsen and M.~Ninomiya, Nucl.\ Phys.\ {\bf
B185} (1981) 20; Nucl\ Phys.\ {\bf B195} (1982) E541; Nucl\ Phys.\
{\bf B193} (1981) 173.
\bibitem{ginsparg82} P.H.~Ginsparg and K.G.~Wilson, Phys.\ Rev.\ {\bf
D25} (1982) 2649.
\bibitem{kaplan92} D.B.~Kaplan, Phys.\ Lett.\ {\bf B288} (1992) 342.
\bibitem{shamir93} Y.~Shamir, Nucl.\ Phys.\ {\bf B406} (1993) 90.
\bibitem{wu99} Nucl.\ Phys.\ (Proc.\ Suppl.) {\bf B83} (2000) 224.
\bibitem{narayanan93} R.~Narayanan and H.~Neuberger, Phys.\ Lett.\
{\bf B302} (1993) 62; Nucl.\ Phys.\ {\bf B412} (1994) 574.
\bibitem{narayanan95} R.~Narayanan and H.~Neuberger, Nucl.\ Phys.\
{\bf B443} (1995) 305.
\bibitem{edwards00} R.G.~Edwards and U.M.~Heller, hep-lat/0005002.
\bibitem{symanzik83} K.~Symanzik, Nucl.\ Phys.\ {\bf B226} (1983) 187;
Nucl.\ Phys.\ {\bf B226} (1983) 205.
\bibitem{SWaction} B.\ Sheikholeslami and R.\ Wohlert, Nucl.\ Phys.\
{\bf B259} (1985) 572.
\bibitem{heatlie91} G.~Heatlie \textit{et al.}, Nucl.\ Phys.\ {\bf
B352} (1991) 266.
\bibitem{hyperfine} C.R.~Allton \textit{et al.} (UKQCD Collaboration),
  Phys.\ Lett.\ {\bf B292} (1992) 408.
\bibitem{lm93} G.P.~Lepage and P.B.~Mackenzie, Phys.\ Rev.\ {\bfseries
D48} (1993) 2250.
\bibitem{alpha96} K.~Jansen \textit{et al.}, Phys.\ Lett.\ {\bf B372}
(1996) 275; 
\bibitem{luscher96} M.~L\"{u}scher \textit{et al.}, Nucl.\ Phys.\ {\bfseries
B478} (1996) 365.
\bibitem{alpha97} M.~L\"{u}scher \textit{et al.}, Nucl.\ Phys.\
{\bfseries B491} (1997) 323.
\bibitem{sommer94} R.~Sommer, Nucl.\ Phys.\ {\bf B411} (1994) 839.
\bibitem{ukqcd94} H.~Wittig (UKQCD Collaboration), Nucl.\ Phys.\
(Proc.\ Suppl.) {\bf 42} (1995) 288.
\bibitem{ukqcd98} C.R.~Allton \textit{et al.} (UKQCD Collaboration),
Phys.\ Rev.\ {\bf D60} (1999) 034507.
\bibitem{philipsen98} O.~Philipsen and H.~Wittig, Phys.\ Rev.\ Lett.\
{\bf 81} (1998) 4056.
\bibitem{knechtli98} F.~Knechtli and R.~Wommer (Alpha Collaboration),
Phys.\ Lett.\ {\bf B440} (1998) 345.
\bibitem{stephenson99} P.W.~Stephenson, Nucl.\ Phys.\ {\bf B550}
(1999) 427.
\bibitem{pennanen00} P.~Pennanen and C.~Michael, hep-lat/0001015
\bibitem{cppacs2000} CP-PACS Collaboration (S.~Aoki \textit{et al}),
Phys.\ Rev.\ Lett.\ 84 (2000) 238.
\bibitem{ukqcd2000} UKQCD Collaboration (K.C.~Bowler \textit{et al}),
{\tt hep-lat/9910022}, Phys.\ Rev.\ {\bf D} (to appear).
\bibitem{sint97} S.\ Sint and P.\ Weisz, Nucl.\ Phys.\ {\bf B502}
(1997) 251.
\bibitem{edwards97} R.G.~Edwards, U.M.~Heller and T.R.~Klassen, Phys.\
Rev.\ Lett.\ 80 (1998) 3448.
\bibitem{michael97} C.~Michael, {\tt hep-ph/9710249}, in Proc.\ of
Advanced Study Institute on Confinement, Duality and Non-Perturbative
Aspects of QCD, Cambridge, England, 23rd.\ June to 4th.\ July, 1997.
\bibitem{morningstar95} C.~Morningstar and M.~Peardon, Nucl.\ Phys.\
(Proc.\ Suppl.) {\bf 42} (1995) 258.
\bibitem{morningstar96} C.~Morningstar and M.~Peardon, Nucl.\ Phys.\
(Proc.\ Suppl.) {\bf 53} (1997) 917; Phys.\ Rev.\ {\bf D56} (1997) 4043.
\bibitem{morningstar99} C.~Morningstar and M.~Peardon, Phys.\ Rev.\ {\bf
D60} (1999) 34509
\bibitem{E852:97} E852 Collaboration (D.R.~Thompson \textit{et al.}A Phys.\
Rev.\ Lett.\ {\bf 79} (1997) 1630; Phys.\ Rev.\ Lett.\ {\bf 81} (1998)
5760.
\bibitem{VES93} G.M.~Beliadze \textit{et al.}, Phys.\ Lett.\ {\bf
B313} (1993) 276; A.~Zaitsev, Proc.\ of HADRON-97, Brookhaven National
Laboratory, August 1997.
\bibitem{lacock96} P.~Lacock, C.~Michael, P.~Boyle and P.~Rowland
(UKQCD Collaboration), Phys.\ Rev.\ {\bf D54} (1996) 
\bibitem{milc98} MILC Collaboration (C.~Bernard \textit{et al.}),
Nucl.\ Phys.\ (Proc.\ Suppl.) {\bf 73} (1999) 264.
\bibitem{milc97} C.~Bernard \textit{et al.} Nucl.\ Phys.\ (Proc.\
Suppl.) {\bf 53} 228.
\bibitem{lacock98} P.~Lacock and K.~Schilling (SESAM Collaboration),
Nucl.\ Phys.\ (Proc.\ Suppl.) {\bf 73} (1999) 261.
\bibitem{lee99} F.X.~Lee and D.B.~Leinweber, Nucl.\ Phys.\ (Proc.\
Suppl.) {\bf B73} (1999) 258.
\bibitem{hamber84} H.~Hamber and C.M.~Wu, Phys.\ Lett.\ {\bf B133}
(1983) 351; Phys.\ Lett.\ {\bf B136} (1984) 255.
\bibitem{riken00} S.~Sasaki, invited talk at Jefferson Laboratory
Workshop on the Physics of Excited Nucleons, Feb.\ 16-19, 2000,
hep-ph/0004252.
\bibitem{capitani98} Nucl.\ Phys.\ (Proc.\ Suppl.) {\bf 73} (1999) 294.
\bibitem{cristov96} C.V.~Christov \textit{et al.}, Prog.\ Part.\
Nucl.\ Phys.\ {\bf 37} (1996) 91.
\bibitem{maiani90} L.~Maiani and M.~Testa, Phys.\ Lett.\ {\bf B245}
(1990) 585.
\bibitem{beccarini95} G.~Beccarini \textit{et al.}, Nucl.\  Phys.\
{\bf B456} (1995) 271.
\bibitem{gockeler95} M.~G\"{o}ckeler \textit{et al}, Phys.\ Rev.\ {\bf
D53} (1996) 2317; Prog.\ Theor.\ Phys.\ Suppl.\ 122 (1996) 145; Phys.\
Rev.\ {\bf D54} (1996) 5705.
\bibitem{cteq95} H.L.~Lai, J.~Botts, J.~Huston, J.G.~Morfin, J.F.~Owens,
J.W.~Qiu, W.K.~Tung and H.~Weerts, Phys.\ Rev.\ {\bf D51} (1995) 4763.
\bibitem{brower96} R.~Brower \textit{et al.}, Nucl.\ Phys.\ (Proc.\
Suppl.) {\bf 53} (1997) 318.
\bibitem{dolgov98} D.~Dolgov \textit{et al.}, Nucl.\ Phys.\ (Proc.\
Suppl.) {\bf 73} (1999) 300.
\bibitem{gockeler97} M.~G\"{o}ckeler \textit{et al.}, Minireview given
at DIS97, Chicago, April 1997, hep-ph/9706502.
\bibitem{capitani99} S.~Capitani \textit{et al.}, Nucl.\ Phys.\
(Proc.\ Suppl.) {\bf 79} (1999) 173; Nucl.\ Phys.\ {\bf B570} (2000) 393.
\bibitem{Lue86} M.~L\"uscher,
Commun.\ Math.\ Phys.\ 105 (1986) 153.
\bibitem{Lue91a}
M~L\"uscher, Nucl.\ Phys.\ {\bf B354} (1991) 531.
\bibitem{Lue91b}
M.~L\"uscher, Nucl.\ Phys.\ {\bf B364} (1991) 237.
\bibitem{Goe94} M.~G\"ockeler, H.A.\ Kastrup, J.\ Westfalen and F.\
Zimmermann, Nucl.\ Phys.\ {\bf B425}
(1994) 413.
\bibitem{Fuk95} M.~Fukugita, Y.~Kuramashi, M.~Okawa, H.~Mino and
A.~Ukawa, Phys.\ Rev.\ {\bf D52} (1995) 3003.
\bibitem{Fuk99} S.~Aoki \textit{et al.} (JLQCD Collaboration),
Nucl.\ Phys.\ (Proc.\ Suppl.) {\bf B83} (2000) 241.
hep-lat/9911025, to appear in the Proceedings of Lattice 99, Pisa, Italy.
\bibitem{Fie00a}
H.R.~Fiebig, H.~Markum, K.~Rabitsch, and A.~Mih\'{a}ly, Few Body
Systems (Suppl.) 10 (1999) 467.
\bibitem{Ric89}
D.G.~Richards, Nucl.\ Phys.\ (Proc. Suppl.) {\bf B9} (1989) 181.
\bibitem{Ric90} D.G.~Richards, D.K.~Sinclair, and D.~Sivers, Phys.\
Rev.\ {\bf D42} (1990) 3191.
\bibitem{Mih97}
A.~Mih\'{a}ly, H.R.~Fiebig, H.~Markum and K.~Rabitsch, Phys.\ Rev.\
{\bf D55} (1997) 3077.
\bibitem{Mic99} C.~Michael, and P.~Pennanen (UKQCD Collaboration),
Phys.\ Rev.\ {\bf D60} (1999) 054012.

\end{thebibliography}
\end{document}